\documentclass[a4paper,11pt]{article}
\usepackage{graphicx,amssymb,amstext,amsmath}
\usepackage{floatflt}
\usepackage{tikz,pgfplots}
\pgfplotsset{compat=1.3}
\usetikzlibrary{spy}

\usepackage{cite}

\definecolor{cbl}{rgb}{0,0,1}                

\usepackage{mathtools}

\newcommand{\bc}{\begin{center}}
\newcommand{\ec}{\end{center}}
\def\ba#1{\begin{array}{#1}\displaystyle}
\newcommand{\ea}{\end{array}}

\newcommand{\beq}{\begin{equation}}
\newcommand{\eeq}{\end{equation}}
\newcommand{\beqa}{\begin{eqnarray}}
\newcommand{\eeqa}{\end{eqnarray}}

\newcommand{\bi}{\begin{itemize}}
\newcommand{\ei}{\end{itemize}}

\newcommand{\bra}{\langle}
\newcommand{\ket}{\rangle}

\newcommand{\Tr}{{\rm Tr}}

\newcommand{\TT}{{\cal T}}

\begin{document}
\begin{titlepage}
\vspace{0.2cm}
\begin{center}
\large{\bf{Entanglement Dynamics after a Quench in Ising Field Theory: \\ A Branch Point Twist Field Approach}}

\vspace{0.8cm} 
{\large \text{Olalla A. Castro-Alvaredo${}^{\heartsuit}$, M\'at\'e Lencs\'es${}^{\diamondsuit}$, Istv\'an M. Sz\'ecs\'enyi${}^{\spadesuit}$, and Jacopo Viti${}^{\clubsuit}$}}

\vspace{0.8cm}
{\small
{\small ${}^{\heartsuit\,\spadesuit}$} Department of Mathematics, City, University of London, 10 Northampton Square EC1V 0HB, UK\\
\vspace{0.2cm}
{ ${}^{\diamondsuit\, \clubsuit}$}International Institute of Physics, UFRN, Campos Universit\'ario, Lagoa Nova 59078-970 Natal, Brazil\\
\vspace{0.2cm}
{ ${}^\clubsuit$}ECT, UFRN, Campos Universit\'ario, Lagoa Nova 59078-970 Natal, Brazil}\\
\end{center}
\medskip
\medskip
\medskip
\medskip

We extend the branch point twist field approach for the calculation of  entanglement entropies to time-dependent problems in 1+1-dimensional massive quantum field theories. We focus on the simplest example:   a mass quench in the Ising field theory from initial mass $m_0$ to final mass  $m$. The main analytical results are obtained from a perturbative expansion of the twist field one-point function in the post-quench quasi-particle basis. The expected linear growth of the R\'enyi  entropies at large times $mt\gg 1$ emerges from a perturbative calculation at second order. We also show that the R\'enyi and von Neumann entropies, in infinite volume,  contain  subleading  oscillatory contributions of frequency $2m$  and amplitude proportional to $(mt)^{-3/2}$. The oscillatory terms are  correctly predicted by an alternative  perturbation  series, in the pre-quench quasi-particle basis,  which we also  discuss. 
A comparison to lattice numerical calculations  carried out on an Ising chain in the scaling limit shows very good agreement with the quantum field theory predictions. We also find evidence of clustering of twist field correlators which implies that the entanglement entropies are proportional to the number of subsystem boundary points.
\vspace{1cm}

\medskip

\noindent {\bfseries Keywords:}  Entanglement Entropy, Quench Dynamics, Ising Model, Integrability,  Branch Point Twist Fields, Perturbation Theory
\vfill

\noindent 
${}^{\heartsuit}$ o.castro-alvaredo@city.ac.uk\\
${}^{\diamondsuit}$ mate.lencses@gmail.com\\
${}^{\spadesuit}$ istvan.szecsenyi@city.ac.uk\\
${}^{\clubsuit}$ viti.jacopo@gmail.com\\

\hfill \today

\end{titlepage}
\section{Introduction}
\label{sec:intro}

Out-of-equilibrium many-body quantum dynamics is one of the most active and challenging research areas in low-dimensional physics; see the special issue~\cite{specialissue2} and in particular the review~\cite{Essler2016}. 
A typical setup triggering out-of-equilibrium evolution from an initial equilibrium state is that of a {\it quantum quench} \cite{quench}. In a quench protocol, a quantum system is prepared at $t<0$ in the ground state, denoted by $|0\ket$, of a Hamiltonian $H(\lambda_0)$ which depends on a parameter $\lambda_0$. At $t=0$, the parameter $\lambda_0$ is suddently changed to a new value $\lambda\not=\lambda_0$ and the unitary time evolution for positive times is governed by the new  Hamiltonian $H(\lambda)$.  The state of the system at time $t$ may be then formally written as $e^{-i t H(\lambda)}|0\ket$.

{\begin{floatingfigure}[h]{7.6cm} 
 \begin{center} 
 \includegraphics[width=7.1cm]{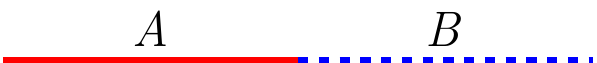} 
 \end{center} 
 \caption{Typical bipartition for the entanglement entropy of two semi-infinite intervals\vspace{1cm}} 
 \label{typical} 
 \end{floatingfigure}}

In this context, the evolution of the bipartite, or von Neumann, entanglement entropy  following a quantum quench has been intensively studied; see~\cite{Calabrese_2009} for a review and references therein. Consider a space bipartition of a 1+1-dimensional quantum system as sketched in fig.~1 and assume that regions $A$ and $B$ are semi-infinite. Then the entanglement entropy associated to region $A$ after a quench may be expressed as $S(t)=-\Tr_A(\rho_A \log \rho_A) $ where formally
\medskip
\beq
\rho_A:=\Tr_B(e^{-i t H(\lambda)}|0\ket \langle 0| e^{it H(\lambda)})\,,
\eeq
 is the reduced density matrix associated to subsystem $A$. Since the regions are semi-infinite, the entropies will not explicitly depend on the subsystem's length. Another set of entanglement measures is provided by the R\'enyi entropies which are defined as
 \beq 
 S_n(t):=\frac{\log \Tr\rho_A^n }{1-n}\,,
 \label{renyi_def}
 \eeq 
and have the property $\lim_{n\rightarrow 1} S_n(t)=S(t)$. It is in fact these R\'enyi entropies which we will mostly be studying in this manuscript.

The universal features of the evolution of entanglement after a quench at a critical point described by conformal field theory have been studied in \cite{EEquench,quench,quench2}. In these works an intuitive picture was put forward, namely one based on the production of highly entangled quasi-particle pairs of opposite momenta right after the quench. These then propagate in space-time until a critical time $t_{\text{sat}}=\frac{\ell}{2 v}$, where $\ell$ is the size of the subsystem and $v$ is the propagation velocity. In this region, the entanglement entropy grows linearly in time. 
For $t_{\text{sat}}>\frac{\ell}{2 v}$ the entanglement saturates to a value proportional to the subsystem's size $\ell$. These features were later demonstrated analytically for the XY chain in a transverse magnetic field in~\cite{FC}, where the exact coefficients of the terms linear in $t$ and in $\ell$ were computed. Note however that for our configuration of two semi-infinite regions $t_{\text{sat}}\rightarrow\infty$. Thus we expect the entanglement to continue to grow linearly in time {\it{for all times}}; this must be beared in mind when comparing analytic results with lattice numerical calculations.

Although the entanglement evolution after a quench has been studied for many physical lattice models, both free \cite{FC,CEF, CEF2, FE,EEquench,quenchesCC,EP,NR,CTC,BFSED,CHMM,BTC, NRV_2018, Bastianello_2018, BFPC_2018} and interacting \cite{Moca_2017, AC2, Mesty_n_2017, ABF_scipost},  complete analytic derivations of such  linear behaviour remained elusive so-far. In $1+1$ dimensions,  they are still  based either on conformal field theory techniques~\cite{EEquench, Asplund2015} or conjectural large space-time asymptotics for block T\"{o}plitz matrices~\cite{FC,CEF,CEF2}. An exception is represented by  models with random unitary evolution such as random circuits~\cite{Nahum_2016, BKP_2019}.

More recently, it has been shown that for gapped systems, the entanglement dynamics after a quench features other  non-trivial effects.  In particular the time-dependence of the entanglement entropy can show subleading corrections  that might qualitatively alter the leading linear increase at large times. For instance, the studies \cite{Gabor1, Gabor2, Gabor3} have shown that  quasi-particle confinement in a linear potential can lead to oscillatory behaviour in time, as well as suppression of linear growth for sufficiently large times.

The purpose of this paper is then twofold: first we will provide a general quantum field theory framework to analyse entanglement dynamics in massive systems. Secondly, we will  provide evidence that subleading oscillatory terms are actually a common feature of entanglement dynamics in infinite volume. 
To this end, we will focus on one of the simplest and best known theories: the Ising field theory.
 We may regard this as the scaling limit of the Ising chain described by the Hamiltonian 
\beq
H_{\text{Ising}}(h)=-J\sum_{i=1}^N \left(\sigma_i^x \sigma_{i+1}^x+h \sigma_i^z\right)\,,
\label{Isingh}
\eeq
where $J>0$, and $h$ is known as the transverse field. The Ising spin chain has a quantum critical point, with a gapless spectrum, at $h=1$, which separates a paramagnetic phase $(h>1)$ from a ferromagnetic phase ($h<1$). The two phases are  related by a Kramers-Wannier duality transformation, which interchanges the spin with the disorder field.

Near the critical point, for $|h-1|\ll 1$ it is possible to define the scaling limit by taking $J\rightarrow \infty$ and $a\rightarrow 0$, where $a$ is the lattice spacing, while keeping 
\beq
m:=2J|h-1|\,,\quad v:=2J a\,,
\label{scalinglimit}
\eeq
fixed and finite~\cite{SE}. In the scaling limit, the low energy excitations of (\ref{Isingh}) are then relativistic real non-interacting fermions with positive mass $m$, while the speed of light is fixed to $v$. 

Finally, a word is due on the techniques that we will be using in this paper. We will exploit the well-known relationship between R\'enyi entropies and correlation functions 
of branch point twist fields \cite{Calabrese:2004eu,Calabrese:2005in,entropy}. For the simple configuration of fig.~1 this means that we will be computing a one-point function  of a branch point twist field $\mathcal{T}(x,t)$ and studying its time dependence after the quench. Branch point twist fields are defined on a replicated quantum field theory containing $n$ identical copies of the original theory, and have been interpreted as symmetry fields associated to the cyclic permutation of the copies in \cite{entropy}. Explicitly, the R\'enyi entropies at time $t$ are given by
\beq
S_n(t)=\frac{\log \left( \varepsilon^{\Delta_n} {}_n\bra 0| \TT(0,t)| 0 \ket_n\right)}{1-n}\,,
\label{entro}
\eeq
 where $\varepsilon$ is some UV cut-off; $\TT(0,t):=e^{i H(\lambda) t} \TT(0,0) e^{-i H(\lambda) t}$ is the time-evolved twist field in the Heisenberg picture; $n$ is the replica number; $|0\ket_n$ is the ground state of the replica theory before the quench that is, for coupling constant $\lambda_0$ (corresponding to mass gap $m_0$);  $\Delta_n$ is the scaling dimension of the twist field at criticality.  For $t=0$, \eqref{entro} gives the R\'enyi entropies at equilibrium in terms of 
\beq
\tau_n:={}_n\bra 0|\TT(0,0)|0\ket_n\,,
\label{taun}
\eeq
 the Vacuum Expectation Value (VEV) of the branch point twist field, which by dimensional analysis  must be proportional to $m_0^{\Delta_n}$. In particular, the UV cut-off $\varepsilon$ is chosen in such a way that no finite $O(1)$ term appears at $t=0$ on the right hand side of~\eqref{entro}.

When comparing our field theoretical predictions for (\ref{entro}) with lattice calculations in the Ising chain in the scaling limit, we will be actually comparing~\eqref{entro} with a similar quantity involving a two-point function of branch point twist fields ${}_n\bra 0| \TT(0,t) \TT^{\dagger} (\ell,t)|0\ket_n$: that is the entanglement entropy of an interval of length $\ell$. To allow for comparison, we will take the length of such interval to be very large, in which case we expect clustering of the two-point function to occur, namely, the factorization 
\beq
\lim_{\ell\rightarrow \infty}{}_n\bra 0| \TT(0,t) \TT^{\dagger} (\ell,t)|0\ket_n \sim {}_n\bra 0| \TT(0,t)|0\ket_n^2\,.
\label{cluster}
\eeq
 Thus our results for (\ref{entro}) will generally give half the values obtained from computations involving a large but finite interval $\ell$. We indeed confirm this in section~\ref{sec:num} of this paper.

\medskip
This paper is organized as follows: in section \ref{sec:sum} we present a summary of our analytical results. In section \ref{sec:tech} we review the  field theoretical tools that we have used:  a time-dependent  formulation of the branch-point twist field approach~\cite{entropy} for the calculation of entanglement measures. In particular, we present an expansion of the twist field one-point function in the post-quench quasi-particle basis following the route traced in~\cite{SE} for the order parameter. In section~\ref{massquench},  we derive  the main analytical results. In section~\ref{pert_theory} we further generalize the perturbative approach to the quench dynamics put forward in~\cite{PQ1} to the calculation of the twist field one-point function. We show that for sufficiently small quenches, these results are in agreement with the main outcome of section~\ref{massquench}. In section~\ref{sec:num} we present a detailed test of our field theoretical predictions against lattice results obtained in the scaling limit.
Finally, we conclude in section~\ref{sec:conc}. An appendix with additional numerical lattice results completes the paper.
\section{Summary of the Main Results}
\label{sec:sum}
Consider the Ising field theory with mass scale $m_0$ and a quench that changes it to a new value $m$. Let us also introduce the function ($\theta\in\mathbb R$)
\begin{equation}
K(\theta)=i\tan\left[\frac{1}{2}\tan^{-1}(\sinh\theta)-\frac{1}{2}\tan^{-1}\left(\frac{m}{m_0}\sinh\theta\right)\right]:=i\hat{K}(\theta)\,,
\label{KK}
\end{equation}
whose meaning we discuss in section~\ref{sec:tech}. Long time after the quench, namely for $mt \gg 1$ the expectation value of the branch point twist field is conjectured to be
{
\beq
{{}_n \bra 0| \TT(0,t) | 0 \ket_n}= \tilde{\tau}'_n  \exp\left[{- \frac{n\Gamma' m t}{2}- \frac{n \mu^2}{64\pi mt}- \frac{\mu}{8 \sqrt{\pi} n} \frac{
\cos\frac{\pi}{2n}}{\sin^2\frac{\pi}{2n}} \frac{\cos(2mt-\frac{\pi}{4})}{(mt)^{\frac{3}{2}}}}+\cdots\right]\,,
\label{twist_exp}
\eeq}
where
\beq
\label{mu}
\mu:= 1-\frac{m}{m_0}=-\frac{\delta m}{m_0}\,, \quad \mathrm{with}\quad \delta m:=m-m_0~.
\eeq
The ellipsis in~\eqref{twist_exp} denote terms that are subleading with respect to $t^{-3/2}$ for large times. The parameters $\tilde{\tau}_n'$ and $\Gamma'$ in \eqref{twist_exp} are calculated perturbatively in the function $K(\theta)$ whose absolute value is then assumed small for $\theta$ real. In particular
\begin{equation}
\label{pert_K}
\tilde{\tau}'_n=\tilde{\tau}_n\, e^{A+O(K^3)}\qquad \mathrm{and} \qquad \Gamma'=\Gamma+O(K^4)\,,
\end{equation}
where $\tilde{\tau}_n$ is the expectation value of the branch point twist field in the post-quench ground state $|\tilde{0}\ket_n$, similar to the definition (\ref{taun}) but with mass gap $m$. The decay rate $\Gamma$ and the constant $A$  in \eqref{pert_K} are
\beq
\Gamma:=2\int_0^\infty \frac{d\theta}{\pi} \, \hat{K}^2(\theta) \sinh \theta\,,
\label{gamma}
\eeq
and
\begin{equation}
\label{A_const}
A:=\frac{1}{2\sin\frac{\pi}{n}}\int_{-\infty}^{\infty}\frac{d\theta}{2\pi}\hat{K}^2(\theta)\,.
\end{equation}
We will provide a full derivation of the twist field one-point function up to $O(K^2)$ and conjecture its general form in~\eqref{twist_exp}  following an analogous calculation as for the one-point function of the spin operator after a mass quench~\cite{SE} in Ising field theory. In particular, we will show that the decay rate $\Gamma$ in \eqref{gamma}  is the same as for the spin operator~\cite{SE}  up to the second-order corrections in the function $K$. The oscillatory contribution in~\eqref{twist_exp} has also the same frequency and power law in $mt$ as for the spin operator, albeit with a different $n$-dependent overall coefficient.

From the explicit calculation of the one-point function of the branch-point twist field, we  can derive an exact expression at $O(K^2)$ for the  R\'enyi entropies (\ref{entro}) at large times after the quench which is given by
\beq
 S_n(t) =\frac{\log(\varepsilon^{\Delta_n}\tilde{\tau}'_n)}{1-n}+ \frac{\Gamma n mt}{2(n-1)}+ \frac{n \mu^2}{64\pi mt (n-1)}+ \frac{\mu}{8 \sqrt{\pi} n} \frac{
\cos\frac{\pi}{2n}}{\sin^2\frac{\pi}{2n}} \frac{\cos(2mt-\frac{\pi}{4})}{(n-1)(mt)^{\frac{3}{2}}}+O(t^{-3})\,.
\label{mainres}
\eeq
Since for $|\mu|\ll 1$, $\hat{K}(\theta)$ is $O(\mu)$ and 
\beq
\Gamma=\frac{\mu^2}{3\pi}+O(\mu^3)\,, \qquad A=\frac{\mu^2}{24 \pi \sin \frac{\pi}{n}}+O(\mu^3)\,.
\label{expgamma}
\eeq
Eq.~\eqref{mainres} can also be viewed  as a large-time expansion  of a perturbative series in the quench parameter $\mu$. In particular, due to \eqref{expgamma}, eq.~\eqref{mainres} is exact up to second-order  terms in $\mu$. This claim will be checked explicitly against analytical and numerical lattice results for the R\'enyi entropies in the scaling limit in section~\ref{sec:num}.

Furthermore, notice that  the oscillating term is $O(\mu)$. Indeed, under the assumption $|\mu|\ll 1$ and provided the replacement $m\rightarrow m_0$ in the frequency, it can be also derived within a first-order perturbative approach to the quench dynamics~\cite{PQ1, PQ2}.
We will postpone details of this alternative derivation to  section~\ref{pert_theory}.  Remarkably, this first-order perturbative approach is not enough to capture 
the leading large-time asymptotics of the R\'enyi entropies. The latter is governed by the decay rate $\Gamma$ and is therefore an $O(\mu^2)$ effect.

Another interesting feature of~\eqref{mainres} is that  the limit $n\rightarrow 1$  is only well-defined for the oscillatory term; for the von Neumann entropy we obtain in particular
\beq
\lim_{n\rightarrow 1} \frac{\mu}{8 \sqrt{\pi} n (n-1)}  \frac{
\cos\frac{\pi}{2n}}{\sin^2\frac{\pi}{2n}} \frac{\cos(2mt-\frac{\pi}{4})}{(mt)^{\frac{3}{2}}}=
  \frac{\mu}{4} \frac{ \sqrt{\pi} \cos(2mt-\frac{\pi}{4}) }{4 (mt)^{3/2}}\,.
 \label{oscEE}
\eeq
 The same limit is obviously ill-defined for all the other contributions in (\ref{mainres}). The reason for this has to do with the way the $O(\mu^2)$ terms are computed, namely from branch point twist fields which are a priori only defined for $n\ \in \mathbb{N}\setminus \{0, 1\}$. Taking the limit generally requires an understanding of the analytic continuation to $n\in \mathbb{R}$ of the one-point function, which is non-trivial, particularly for higher particle form factor contributions (e.g. precisely the ones that give rise to the problematic terms in (\ref{mainres})). We have not carried out this limit here, but we have found good agreement between (\ref{oscEE}) and lattice calculation in the scaling limit for the Neumann entropy. We report these comparisons in section~\ref{sec:num}

\section{Review of the Main Techniques}
\label{sec:tech}
In this section we  review the  main techniques that we have employed in order to derive the results of the previous section: branch point twist fields in relation to entanglement measures in the Ising field theory and the expansion of the one-point function of a local operator in the post-quench quasi-particle basis, as developed in~\cite{SE}. Results obtained from a perturbative approach~\cite{PQ1,PQ2} in the quench parameter will be given in section~\ref{pert_theory}.

\subsection{Branch Point Twist Fields and Entanglement}
\label{TField}
The main properties of branch point twist fields were described at length in \cite{entropy} and the subsequent review article \cite{review}. 
 In  quantum field theory, it has been known for some time \cite{Calabrese:2004eu, Calabrese:2005in, entropy} that, for integer $n$, the R\'enyi entropies in \eqref{renyi_def}  may be expressed in terms of correlation functions of branch point twist fields, with the number of twist field insertions equalling the number of boundary points of the subsystems under consideration. This means that the entanglement entropy of a semi-infinite region is simply given by the one-point function of the branch point twist field.
Eq.~\eqref{entro} can be thought as the obvious time-dependent generalization of the setting in~\cite{entropy} at equilibrium.
 
Before embarking into the study of the time-dependent one-point function of the twist field,  it is useful to recall some of its properties at equilibrium.
 At equilibrium, the $n^{\mathrm{th}}$ R\'enyi entropy of a semi-infinite system in the ground state of $H(\lambda_0)$ is given by~(\ref{entro}) at $t=0$, namely
 \beq
S_n(0)=\frac{\log\left(\varepsilon^{\Delta_n}\tau_n \right)}{1-n}\,,
\label{renyi}
\eeq
Obviously in the ground  state $|\tilde{0}\ket_n$ of the post-quench Hamiltonian $H({\lambda})$ (with a  mass gap $m$)~\eqref{renyi} applies by replacing $\tau_{n}\rightarrow\tilde{\tau}_n$.
The power $\Delta_n$ is 
 the scaling dimension of the branch point twist field at criticality, which is given by
\beq
\label{stwist}
\Delta_n=\frac{c}{12}\left(n-\frac{1}{n}\right)\,,
\eeq
where $c$ is the central charge of the underlying CFT \cite{kniz,Bouwknegt, Calabrese:2004eu};  for instance,  $c=\frac{1}{2}$ for Ising field theory. 
The parameter $n$, which was already introduced in section~\ref{sec:intro}, is the number of copies of the replicated Hilbert space of the quantum field theory, upon which the branch point twist field acts. Therefore, in such a replicated theory, the pre- and post-quench ground states $|0\ket_n$ and $|\tilde{0}\ket_n$  are tensor products of $n$ copies of the physical ground states defined in the introduction. The same construction carries over for the time-dependent case.

 Finally, it is useful to remember that when comparing with lattice results for the Ising spin chain, the natural choice for the cut-off is the lattice spacing $a$. The UV cut-off $\varepsilon$ and $a$ are related by a model-dependent (i.e.~non-universal)  proportionality constant. Therefore on the lattice \eqref{renyi} reads~\cite{Calabrese:2004eu}
\begin{equation}
\label{scal_eq}
 S_n(0)=-\frac{c}{12}\left(n+\frac{1}{n}\right)\log(m_0 a)+ O(1)\,,
\end{equation}
where $O(1)$ denotes non-universal terms that are finite or vanish in the scaling limit $a\rightarrow 0$.
 The leading logarithmic lattice spacing dependence in~\eqref{scal_eq} can be used to extract the twist field scaling dimension, \textit{alias} the central charge of the UV fixed point, in lattice numerical calculations. Actually the quality of such an extrapolation provides  a useful measure of how close the numerical calculation is to the scaling regime of the lattice model; see  section~\ref{sec:num} and appendix \ref{app:a}. 

\subsection{Expansion of the Time-Dependent One-Point Function in the Post-Quench Basis}
\label{sec:formfactor}

In this section we review the approach first employed  in \cite{SE} to study relaxation dynamics of a local operator in the Ising field theory after a mass quench. The technique needs two inputs: an expansion of the initial state into eigenstates of the post-quench Hamiltonian and the matrix elements of the local operator one is interested in, between states of the post-quench quasi-particle basis. In principle the method is applicable  also to interacting post-quench theories, provided that such analytical data are known; in particular the post-quench theory, considered in infinite volume and for all times, 
should be integrable. See for instance~\cite{Hodsagi19} for recent activity devoted to overlap calculations. 

In the specific case of the Ising mass quench, the non-normalized initial state $|\Omega\rangle:=\sqrt{\langle\Omega|\Omega\rangle}|0\rangle$, $|0\rangle$ being the ground state of the pre-quench Hamiltonian, can be \textit{exactly} expressed in terms of eigenstates of the post-quench Hamiltonian as
 \beq
 |\Omega\ket=\exp\left[\int_0^\infty \frac{d\theta}{2\pi} K(\theta) a^\dagger(-\theta) a^\dagger(\theta) \right]|\tilde{0}\ket\,.
 \label{psio}
 \eeq 
Notice therefore that $|\tilde{0}\ket$ is the vacuum of the post quench Ising field theory (i.e. with mass gap $m$); $a^\dagger(\theta)$ is the fermionic creation operator,  and $K(\theta)$ is the function given earlier in (\ref{KK}). The integral in~\eqref{psio} is over the so-called rapidity which parametrizes the energy ($E$) and momentum ($P$)  of the one-particle  state $|\theta\rangle:= a^{\dagger}(\theta)|\tilde{0}\rangle$ as follows: $E=m\cosh\theta$ and $P=m\sinh\theta$. The normalization of the one-particle states is $\langle\theta|\theta'\rangle=2\pi\delta(\theta-\theta')$.

Quenches leading to states with the structure (\ref{psio}) where studied in detail in \cite{MFi,SFM}. A derivation of the function (\ref{KK}) is given in Appendix A of \cite{SFM}. These states have the same structure of the boundary states first described by Ghoshal and Zamolodchikov \cite{Gosh}. Their structure neatly fits with the quasi-particle picture put forward in \cite{EEquench,quench,quench2} as the initial state (\ref{psio}) can be regarded as a coherent superposition of particle pairs, also known as a squeezed coherent state. Exact solvability of the quench dynamics, which is generally not possible, has been also related~\cite{Piroli2017} to initial states analogous to~\eqref{psio}, see for instance \cite{PQ1}.

In the $n$-copy theory, this simply generalises to
\beq
 |\Omega\ket_n=\exp\left[\sum_{j=1}^n\int_0^\infty \frac{d\theta}{2\pi} K(\theta) a_j^\dagger(-\theta) a_j^\dagger(\theta) \right]|\tilde{0}\ket_n\,,
 \label{nstate}
 \eeq 
where $a_j^\dagger(\theta)$ is the fermionic creation operator in copy $j$. We denote by
$|\theta_1,\ldots, \theta_k\ket_{j_1,\ldots,j_k;n}$  an element of an orthonormal basis in the replicated ($in$ or $out$) Hilbert space  consisting of $k$ particles with rapidities $\theta_i$ and copy labels $j_i$, $i=1,\ldots, k$. The energy and momentum of  multi-particle states are  the sum of the energies and momenta of their one-particle constituents. 

In such a framework, the R\'enyi entropies after the quench can be written as
 \begin{equation}
 S_n(t)=\frac{1}{1-n}\log\left(\frac{ \varepsilon^{\Delta_n} {}_n \bra \Omega| \TT(0,t)|\Omega\ket_n}{{}_n \bra \Omega|\Omega\ket_n}\right).
 \label{t_SE}
 \end{equation}
Substituting the representation \eqref{nstate} of the replicated initial state into~\eqref{t_SE}, both numerator and denominator admit a formal expansion as sums of integrals of matrix elements in the post quench basis. Borrowing notations from~\cite{SE}, we will write these series  as 
 \beq
  {}_n \bra \Omega| \TT(0,t)|\Omega\ket_n:=\tilde{\tau}_n\sum_{k_1, k_2=0}^\infty  C_{2k_1, 2k_2}(t)\,,
  \label{1po}
 \eeq
 with
 \begin{align}
&\tilde{\tau}_nC_{2k_1, 2k_2}(t)= \frac{1}{k_1! k_2!} \sum_{j_1,\ldots, j_{k_1}=1}^n  \sum_{p_1,\ldots, p_{k_2}=1}^n  \nonumber\\
& \times \left[\prod_{s=1}^{k_1 }\int_{0}^\infty \frac{d\theta'_{s}}{2\pi}K(\theta'_{s})^* e^{2 i t E(\theta'_{s})}\right] \left[\prod_{r=1}^{k_2 }\int_{0}^\infty \frac{d\theta_{r}}{2\pi}K(\theta_{r})e^{-2  i t E(\theta_r)}\right]\nonumber\\
 & \times  {}_{n;j_1j_1\ldots j_{k_1} j_{k_1}} \bra \theta'_1,-\theta'_1, \ldots, \theta'_{k_1},-\theta'_{k_1}|\TT(0,0)| -\theta_{k_2},\theta_{k_2},\dots, -\theta_{1}, \theta_1 \ket_{p_{k_2}p_{k_2}\ldots p_1 p_1;n}\,,
  \label{C22}
 \end{align}
 and analogously
 \beq
 {}_n \bra \Omega|\Omega\ket_n:=\sum_{k=0}^\infty  Z_{2k}\,,
 \label{zpo}
 \eeq
 where now
  \beqa
&& Z_{2k}= \frac{1}{(k!)^2} \sum_{j_1,\ldots, j_{k}=1}^n  \sum_{p_1,\ldots, p_{k}=1}^n\left[\prod_{s=1}^{k }\int_{0}^\infty \frac{d\theta'_{s} d\theta_s}{(2\pi)^2}K(\theta'_{s})^* K(\theta_{s})\right]\nonumber\\
 && \times  {}_{j_1j_1\ldots j_{k} j_{k}} \bra \theta'_1,-\theta'_1, \ldots, \theta'_{k},-\theta'_{k}| -\theta_{k},\theta_{k},\dots, -\theta_{1}, \theta_1 \ket_{p_{k}p_{k}\ldots p_1 p_1} \quad \mathrm{for}\quad  k>0\,,
 \label{Z22}
 \eeqa
 and $Z_0=1$. The ratio in~\eqref{t_SE} can be then expanded formally in powers of the function $K$ 
 \begin{equation}
 \label{d_series}
 \frac{{}_n \bra \Omega| \TT(0,t)|\Omega\ket_n}{{}_n \bra \Omega|\Omega\ket_n}:=\tilde{\tau}_n\sum_{k_1,k_2=0}^{\infty} D_{2k_1 2k_2}(t)\,,
 \end{equation}
with 
\beq 
D_{2k_1,2k_2}(t)=\sum_{p=0}^{\min(k_1,k_2)} \tilde{Z}_{2p} C_{2(k_1-p),2(k_2-p)}(t)\,, \label{eq:D_2k_2l_original}
\eeq 
where $\tilde{Z}_{2p}$ are the expansion coefficients of the inverse of the norm, i.e. $\sum_{k,p=0}^\infty Z_{2k}\tilde{Z}_{2p}=1$. In section~\ref{massquench} we present the calculation up to $O(K^2)$.
 
 The matrix elements of the twist field in~\eqref{C22}  can be related to the so-called elementary form factors~\cite{KW,SmirnovBook,entropy}, see~\eqref{paf} in the next section.  
 The transformation that relates the two functions  is called crossing. Consider for instance the matrix element $_{n;j_1}\bra \theta_1|\mathcal{T}(0,0)|\theta_2\ket_{j_2;n}$. This can be written as
 \beqa
\label{first}
	_{n;j_1}\bra \theta_1|\mathcal{T}(0,0)|\theta_2\ket_{j_2;n} &=&\tilde{\tau}_n  {\,} _{n;j_1}\bra \theta_1|\theta_2\ket_{j_2;n} 
    +{}_n\bra \tilde{0}|\mathcal{T}(0,0)|\theta_1+i\pi-i\eta,\theta_2\ket_{j_1,j_2;n}\nonumber\\
    &=&2\pi\,\tilde{\tau}_n \, \delta(\theta_{12})\delta_{j_1j_2}+F_2^{j_1 j_2}(\theta_{12}+i\pi-i\eta)\,,
\eeqa
where $\theta_{12}:=\theta_1-\theta_2$, $\eta$ is a small positive parameter and $F_{2}^{j_1 j_2}(\theta)$ will be given in~\eqref{full}. This relation  can be generalized to matrix elements involving states with larger number of particles~\cite{SmirnovBook}. The shift by $i\eta$ makes the function $F_{2}^{j_1 j_2}(\theta)$ on the right hand side of \eqref{first} regular for $\theta\rightarrow i\pi$.  There are however additional sources of divergences related to the normalization of the asymptotic states in infinite volume, see the $\delta$ function in \eqref{first}.

These infinite volume singularities are expected to be cancelled by similar singularities in the denominator in \eqref{Z22} in the combination as \eqref{eq:D_2k_2l_original}. The precise way in which this cancellation occurs has been the object of much investigation over the past decade and a rigorous understanding now exists. That is, to consider the theory in finite volume $V$ and use the volume as regulator \cite{PT1,PT2}. However, this rigorous approach is rather involved and for this reason some simpler methods, have also been developed. In~\cite{SE} a regularization scheme known as  $\kappa$-regularization \cite{EK1,EK2} was used.  The technique requires to shift the coinciding rapidities by a real value $\kappa$  (or several values $\kappa_i$ for multi-particle states) so that the singularities are avoided. Then introduce a smooth function $P(\kappa)$ which is strongly peaked about $\kappa=0$ with the properties
\beq
P(0)=V\,, \qquad \mathrm{and} \qquad \int_{-\infty}^\infty d\kappa \, P(\kappa) =1\,. 
\eeq
Of course, there are many functions that would meet the criteria above but one expects that in the infinite volume limit $V\rightarrow \infty$ they will all lead to the same finite result. A natural choice also employed in \cite{SE} is a gaussian $P(\kappa)=V e^{-\pi \kappa^2 V^2}$. For instance for a two-particle form factor the regularization would  be implemented as
\beq 
_{n;j_1}\bra \theta_1|\mathcal{T}(0,0)|\theta_2\ket_{j_2;n}\mapsto\int_{-\infty}^{\infty} d\kappa\, P(\kappa)~_{n;j_1}\bra \theta_1|\mathcal{T}(0,0)|\theta_2+\kappa\ket_{j_2;n} \,.
\eeq 
After using the crossing relation~\eqref{first}, it is possible to  isolate the infinite volume divergences, coming from the normalization of the states, and the leading contribution for $V\rightarrow\infty$, by expanding the integrand as a series about $\kappa=0$. Applications can be found in~\cite{SE} and in section~\ref{massquench}.

\subsection{Form Factors of the Branch Point Twist Field in  the Ising Field Theory}
\label{sec:ff}

In this  section, we will finally recall the necessary results for the form factors of the twist field in the  Ising field theory. The explicit form of the form factors is needed to evaluate the  numerator of~\eqref{t_SE}. In the replicated Ising field theory  fermionic particles have an extra copy index $j=1,\dots,n$. There is also an internal $\mathbb{Z}_2$ symmetry in each copy which implies, for $\mathbb Z_2$ even fields, such as the twist field, that only even-particle form factors are non-vanishing. As already mentioned in the previous section, let $|\theta_1,\ldots, \theta_k\ket_{j_1,\ldots,j_k;n}$ be an asymptotic \textit{in} state of the replicated theory consisting of $k$ particles with rapidities $\theta_i$ and copy labels $j_i$, $i=1,\ldots, k$.  We further assume
 $\theta_1>\theta_2> \cdots >\theta_k$. The two-particle twist field form factor  is defined as~\cite{entropy}
  \beq
F_2^{j_1 j_2}(\theta_1-\theta_2):= {}_n \bra 0|\TT(0,0)|\theta_1\theta_2\ket_{j_1,j_2; n}\,,
\label{ff2}
\eeq
and is given by
 \beq
F_2^{j_1 j_2}(\theta)=\frac{\tau _n \sin\frac{\pi}{n}}{2n \sinh\left[\frac{i\pi(1+2(j_1-j_2))+ \theta}{2n}\right]\sinh\left[\frac{i\pi(1-2(j_1-j_2)) -\theta}{2n}\right]} \frac{F_{\text{min}}^{j_1j_2}(\theta)}{F_{\text{min}}^{11}(i\pi)}\,,
\label{full}
\eeq
with
\begin{equation}
F_{\text{min}}^{j_1 j_2}(\theta)=\begin{cases}
-i\sinh[\frac{\theta+2\pi i(j_1-j_2)}{2n}] & j_1\geq j_2\\
+i\sinh[\frac{\theta+2\pi i(j_1-j_2)}{2n}] & j_1< j_2
\end{cases}\,,
\label{is}
\end{equation}
and $\tau_n$ was defined in (\ref{taun}).
Here we have used the fact that the twist field is a Lorentz scalar and therefore the form factor depends only on the rapidity difference, rather than two separate rapidities. Due to the free nature of the theory the $k$-particle form factors are given in terms of Pfaffians~\cite{nexttonext},
\beq 
\label{paf}
F_k^{j_1\ldots j_k}(\theta_1, \ldots, \theta_k):= {}_n \bra 0|\TT(0)|\theta_1\ldots \theta_k\ket_{j_1,\ldots, j_k ; n}=\tau_n {\mathrm{Pf}}(W)\,,
\eeq 
where $\mathrm{Pf}$ is the {\it Pfaffian} of the matrix $W$ (i.e. $\mathrm{Pf}^2(W) =\det(W)$), which  in turn is defined as
\beq 
W_{j_i j_r}=\frac{{F_{2}^{j_i j_r}(\theta_i-\theta_r)}}{\tau_n}\,.\label{K}
\eeq 
In practice, \eqref{paf} implies that, if we call the two-particle form   factor in~\eqref{K} a contraction, $k$-particle form factors ($k$ even) are obtained as sums of products of contractions as prescribed by the  Wick theorem for fermionic fields.
Due to the particular monodromy properties of the branch point twist field discussed in \cite{entropy}, all form factors can be ultimately expressed in terms of form factors involving only one copy of the theory. In particular
\beq
\label{shiftf}
F_k^{j_1\ldots j_k}(\theta_1, \ldots, \theta_k)=F_k^{1\ldots 1}(\theta_1+2\pi i (j_1-1), \ldots, \theta_k+2\pi i (j_k-1))\,,
\eeq
for $j_1\geq j_2 \geq \cdots \geq j_k$. For the Ising field theory this means that the two-particle form factor of particles in the first copy is effectively the building block for any other form factor. For this reason it is useful to adopt a simpler notation for this form factor. We then define the normalized two-particle form factor
\beq
f(\theta):=\frac{{F_{2}^{11}(\theta)}}{\tau_n}\,.\label{norma}
\eeq
All formulas presented in this section are valid when considering the post-quench ground state $|\tilde{0}\rangle_n$, if one replaces $\tau_n$ with $\tilde{\tau}_n$.

 \section{R\'enyi Entropies after a Mass Quench: Field Theory Results}
 \label{massquench}
 As outlined at the beginning of section~\ref{sec:formfactor}, the calculation is organized as a perturbation series in  powers of the function $K$ introduced in (\ref{KK}). In principle, the final result is not  limited to $\delta m\ll 1$, see~\eqref{mu}, provided $K(\theta)$ is sufficiently small for $\theta\in\mathbb R$. 
Physically this is equivalent to truncating the series in~\eqref{psio} to a few multi-particle states. In this section we will fill in the details of the derivation of~\eqref{mainres}.
 \subsection{Contributions at $O(K)$}

Apart from  a trivial $K$-independent term, corresponding to $C_{00}=D_{00}=1$ in~\eqref{1po}, the leading term in the $K$ expansion of~\eqref{t_SE} is $O(K)$ and given by 
 \beqa
C_{2,0}(t)+C_{0,2}(t)&=& n\left[ \int_0^\infty \frac{d\theta}{2\pi} K(\theta)^* f(2\theta) e^{2itE(\theta)} +\int_0^\infty \frac{d\theta}{2\pi} K(\theta) f(2\theta)^* e^{-2itE(\theta)}\right] \nonumber\\
&=&
-  \int_0^\infty \frac{d\theta}{2\pi}  \hat{K}(\theta) \frac{2\cos\frac{\pi}{2n}\sinh\frac{\theta}{n}}{\sinh\frac{i\pi-2\theta}{2n} \sinh\frac{i\pi+2\theta}{2n}} \cos\left[2m t\cosh\theta \right]\,,
\label{C2002}
\eeqa
where we have used~\eqref{norma} and~\eqref{full}. Notice that the expansion of the denominator in~\eqref{zpo} starts as $1+O(K^2)$, therefore, see~\eqref{eq:D_2k_2l_original}, $C_{2,0}+C_{0,2}=D_{2,0}+D_{0,2}$.
 At large times, according to stationary phase analysis, we can expand the integrand in~\eqref{C2002} close to $\theta=0$ and observe that 

\beq
\hat{K}(\theta)= \frac{\mu}{2} \theta+O(\theta^3)\,,
\eeq
with $\mu$ defined by (\ref{mu}).
 By retaining only contributions up to  $O(K)$,  the one-point function of the twist field is then for $mt\gg 1$ 
\begin{equation}
\frac{{}_n\bra \Omega |\TT(0,t)|\Omega\ket_n}{{}_n\bra \Omega|\Omega\ket_n}=\tilde{\tau_n}\left(1-\frac{\mu}{8 \sqrt{\pi} n}  \frac{\cos\frac{\pi}{2n}}{\sin^2 \frac{\pi}{2n}} \frac{\cos(2mt-\frac{\pi}{4})}{(mt)^{3/2}}+\dots\,\right)+O(K^2)\,.
\label{mismo2}
\end{equation}
As anticipated in section~\ref{sec:intro} for $|\mu|\ll 1$   the same result can be derived from a perturbation theory approach~\cite{PQ1}; see section~\ref{pert_theory} for details. We will show in a subsequent section that the terms above are in fact just the first two contributions to the expansion of an exponential, hence the expression (\ref{twist_exp}).

\subsection{Contributions at $O(K^2)$}
\label{sec:ksquare}
 The $O(K^2)$ contributions are considerably more involved and provide a first indication that R\'enyi entropies after the quench grow linearly in time.  Taking into account numerator and denominator in \eqref{t_SE}, the $O(K^2)$ contributions in the expansion of the one-point function are   given by, see again~\eqref{eq:D_2k_2l_original}
 \begin{equation}
 \label{d22}
 D_{2,2}(t)=C_{2,2}(t)-Z_2 C_{0,0}\,,
 \end{equation}
 and 
 \begin{equation}
 \label{d40}
  D_{0,4}(t)+D_{0,4}(t)=C_{0,4}(t)+C_{4,0}(t)\,.
 \end{equation}
 
\subsubsection{The Contribution $D_{2,2}$}
 
We start analysig $D_{2,2}$ in \eqref{d22}; from~\eqref{C22}  one has
 \beqa
\tilde{\tau}_n C_{2,2}(t)
&=&\sum_{j,p=1}^n \int_{0}^\infty \frac{d \theta d \theta'}{(2\pi)^2} M(\theta',\theta; t) \,{}_{n;jj}\bra \theta', -\theta'|\TT(0,t)|-\theta+\kappa, \theta+\kappa\ket_{pp;n} \nonumber\\
&=& n\sum_{j=1}^n \int_{0}^\infty \frac{d \theta d \theta'}{(2\pi)^2} M(\theta',\theta; t) \,{}_{n;11}\bra \theta', -\theta'|\TT(0,t)|-\theta+\kappa, \theta+\kappa\ket_{jj;n}\,.
\label{original}
 \eeqa
 The second equality follows from permutation symmetry of the replicas, and we also defined
\beq
\label{mfun}
M(\theta',\theta;t):=\hat{K}(\theta')\hat{K}(\theta)e^{2imt[\cosh\theta'-\cosh\theta]}\,.
\eeq
Finally,  $C_{0,0}=1$ and from~\eqref{zpo} it follows 
\beq
Z_2=n\sum_{j=1}^{n}\int_{0}^{\infty}\frac{\mathrm{d}\theta'}{2\pi}\int_{0}^{\infty}\frac{\mathrm{d}\theta}{2\pi}\hat{K}(\theta')\hat{K}(\theta)\,{}_{n;11}
\negmedspace\left\langle \theta',-\theta'|-\theta+\kappa,\theta+\kappa\right\rangle _{jj;n}\,.
\label{z2t}
\eeq
To further manipulate~\eqref{original}, we exploit the crossing relation~\cite{SmirnovBook}
\begin{multline}
{}_{n;11}\bra \theta', -\theta'|\TT(0,0)|-\theta+\kappa, \theta+\kappa\ket_{jj;n}
=\\(2\pi)^2 \tilde{\tau}_n \left[\delta(\theta'-\theta-\kappa) \delta(-\theta'+\theta-\kappa) \delta_{1j}-
\delta(\theta'+\theta-\kappa) \delta(-\theta'-\theta-\kappa) \delta_{1j}\right]\\
- 2\pi \left[\delta(-\theta'-\theta-\kappa) F_2^{1j}(\theta'_+-i\eta+\theta-\kappa)-\delta(\theta'-\theta-\kappa) F_2^{1j}(-\theta'_{-}-i\eta +\theta-\kappa)\right]\delta_{1j}\\
- 2\pi \left[\delta(\theta'+\theta-\kappa) F_2^{1j}(-\theta'_{-}-i\eta-\theta-\kappa)-\delta(-\theta'+\theta-\kappa) F_2^{1j}(\theta'_+-i\eta -\theta-\kappa)\right]\delta_{1j}\\
+ F_4^{11jj}(\theta'_+-i\eta_1,-\theta'_{-}-i\eta_2,-\theta+\kappa, \theta+\kappa)\,,
\label{bringover}
\end{multline}
which generalises  (\ref{first}) to four-particle states. In~\eqref{bringover} and hereafter, we used the notation: $\theta_{\pm}:=\theta \pm i\pi\,$. After substituting~\eqref{bringover} into~\eqref{original} we  regroup the result into three terms: $C_{2,2}:=C_{2,2}^{(0)}+C_{2,2}^{(2)}+C_{2,2}^{(4)}$. The integrand of 
$C_{2,2}^{(0)}$ contains the first line in~\eqref{bringover}, the integrand of $C_{2,2}^{(2)}$ contains the second and third line in~\eqref{bringover} while the integrand of $C_{2,2}^{(4)}$ includes the four-particle form factor in the last line of~\eqref{bringover}. Now it is easy to see that $C_{2,2}^{(0)}=Z_2C_{00}$ and therefore the only non-vanishing contribution at $O(K^2)$ is, see~\eqref{d22},
\begin{equation}
 \label{c22_imp}
 D_{2,2}(t)=C_{2,2}^{(2)}(t)+C_{2,2}^{(4)}(t)\,.
\end{equation}
The double integral $C_{2,2}^{(2)}$, after integrating the delta function over $\theta'$ and exploiting the symmetries $M(-x,y;t)=-M(x,y;t)$ and $M(-x,-y; t)=M(x,y;t)$, can be rewritten as
\begin{equation}
\label{C_222}
C_{2,2}^{(2)}(t)=n f(-2\kappa+i\pi-i\eta)\int_{-\infty}^{\infty}\frac{d\theta}{2\pi} M(\theta+\kappa,\theta; t)\,.
\end{equation}
Notice that the sum over $j$ in~\eqref{original} reduces in this case to only one term, due to the Kronecker delta in~\eqref{bringover}. In the $\kappa$ regularization scheme, eq.~\eqref{C_222} should be first integrated with the measure $P(\kappa)$, discussed in section~\ref{sec:tech}, and then the outcome  of the integration evaluated in the limit $V\rightarrow\infty$ and $\eta\rightarrow 0$. In practice, one expands~\eqref{C_222} in a power series in $\kappa$ close to $\kappa=-i \eta/2$ and observes that $\int d\kappa P(\kappa)\kappa^n=O(V^{-n})$, therefore in the infinite volume limit only terms that are singular or finite for $\kappa,\eta\rightarrow 0$ contribute to the final result. Actually, when summed up at a given order in $K$, divergent terms in $\kappa$ should cancel consistently. Expanding the function $C_{2,2}^{(2)}$ around $\kappa=-i \eta/2$ we obtain
\begin{multline}
\label{C22_kappa}
C_{2,2}^{(2)}(t)=-\frac{i n}{2\kappa+i\eta}\int_{-\infty}^{\infty}\frac{d\theta}{2\pi}\hat{K}^2(\theta)+\frac{1}{2\sin\frac{\pi}{n}}\int_{-\infty}^{\infty}\frac{d\theta}{2\pi}\hat{K}^2(\theta)\\
-\frac{in}{2}\int_{-\infty}^{\infty}\frac{d\theta}{2\pi}\frac{d\hat{K}(\theta)}{d\theta}\hat{K}(\theta)+nmt\int_{-\infty}^{\infty}\frac{d\theta}{2\pi}\hat{K}^2(\theta)\sinh(\theta)+O(\kappa)\,.
\end{multline}
The third and fourth terms vanish by symmetry, while the first one which is divergent in the limit $\kappa,\eta\rightarrow 0$, will be cancelled by an opposite contribution coming from $C_{2,2}^{(4)}$. In conclusion only the second time-independent term in~\eqref{C22_kappa}, contributes to the final result for the twist field one-point function. Such a  constant was called $A$ in~\eqref{A_const}. 

Let us then finally analyze $C_{2,2}^{(4)}$. This is a double integral weighted by the function $M(\theta',\theta; t)$ of the four-particle form factor in~\eqref{bringover}. For the Ising model such a form factor is obtained, see the definition (\ref{paf}), applying the Wick theorem as
\begin{multline}
\tilde{\tau}_n F_4^{11jj}(\theta'_+-i\eta_1,-\theta'_{-}-i\eta_2,-\theta+\kappa, \theta+\kappa)= F_2^{11}(2\theta'-i(\eta_1-\eta_2))F_2^{jj}(-2\theta)\\
  -F_2^{1j}(\theta'_{+}+\theta -i\eta_1-\kappa)F_2^{1j}(-\theta'_{-}-\theta-i\eta_2-\kappa)\\
 + F_2^{1j}(\theta'_{+}-\theta-i\eta_1-\kappa) F_2^{1j}(-\theta'_{-}+\theta-i\eta_2-\kappa)\,.
 \label{1stlinethis}
\end{multline}
Using~\eqref{shiftf} and in particular $F_{2}^{1j}(\theta)=\tilde{\tau}_n f(2\pi i(j-1)-\theta)=-\tilde{\tau}_n f(\theta-2\pi i(j-1))$ for $j\neq1$, we can rewrite  the two-particle form factors in~\eqref{1stlinethis} in terms of the elementary function $f$, see section~\ref{sec:tech}.  The sums over $j$, needed to construct $C_{2,2}^4$, see~\eqref{original}, can be then performed by using the identity
\begin{align}
G(x,y):=&\sum_{j=1}^{n} f(-x+2\pi i j) f(y+2\pi i j)\nonumber\\
&=-\frac{i \sinh \frac{x+y}{2}}{2 \cosh \frac{x}{2} \cosh \frac{y}{2}}
\left[f(x+y+i\pi)+f(x+y-i\pi) \right]\,,
\label{Gsum}
\end{align}
that can be found for instance in the Appendix of \cite{nexttonext}. This gives
\begin{multline}
\sum_{j=1}^n F_4^{11jj}(\theta'_+-i\eta_1,-\theta'_{-}-i\eta_2,-\theta+\kappa, \theta+\kappa)= n \tilde{\tau}_n f(2\theta'-i(\eta_1-\eta_2))f(-2\theta)\\
 + \tilde{\tau}_n [G(\theta'_+-\theta -i\eta_1-\kappa,\theta'_{-}-\theta+i\eta_2+\kappa)-G(\theta'_++\theta -i\eta_1-\kappa,\theta'_{-}+\theta+i\eta_2+\kappa)]\,. 
 \label{ggg}
 \end{multline}
 The two lines in \eqref{ggg} have to be finally integrated over the rapidities $\theta$ and $\theta'$ to obtain the function $C_{2,2}^4(t)$. We define $I(t)$ to be the result of integrating the second line in \eqref{ggg} (i.e. the function inside the square bracket) and $I'(t)$ to be the result of integrating the first line (i.e. the product of functions $f$). In this way $C_{2,2}^{4}(t)=I(t)+I'(t)$; we start by  analyzing $I'(t)$ which is simply
 \begin{align}
 \label{Iprime}
 I'(t)&=n^2\int_{0}^{\infty}\frac{d\theta d\theta'}{(2\pi)^2}M(\theta',\theta; t)f(2\theta')f(-2\theta)=|C_{2,0}(t)|^2\,,
 \end{align}
 $C_{2,0}$ given in~\eqref{C2002}; the result follows from $f(\theta)=-f(\theta)^*$ for $\theta\in\mathbb R$. Notice that, according to the discussion in~\eqref{oscEE}, for large times $I'(t)=O(t^{-3})$.
 
 The remaining integral to complete our calculation of $D_{2,2}(t)$ is $I(t)$. After substituting the explicit form for the function $G$, given in \eqref{Gsum} into \eqref{ggg} and using $M(x,y;t)=-M(x,-y;t)$ it can be eventually rewritten as
 \begin{eqnarray}\label{inti}
I(t)= n\int_{0}^{\infty}\frac{\mathrm{d}\theta'}{2\pi}\int_{-\infty}^{\infty}\frac{\mathrm{d}\theta}{2\pi}M(\theta',\theta; t)\frac{H(\theta',\theta)}{2\sinh(\frac{\theta'-\theta-\kappa-i\eta_{1}}{2})\sinh(\frac{\theta'-\theta+\kappa+i\eta_{2}}{2})}\,,
\end{eqnarray}
where we have introduced the function
\beq
\label{funH}
H(\theta',\theta)=-i\sinh\left(\theta'-\theta-\frac{i\eta_{12}}{2}\right)\left[f(2\theta'-2\theta-i\eta_{12}+i\pi)+f(2\theta'-2\theta-i\eta_{12}-i\pi)\right]\,,
\eeq
which is regular along the integration contour in the variable $\theta$ in the limit $\eta_{1,2}\to 0$; also $\eta_{12}:=\eta_1-\eta_2$.

 The denominator in~\eqref{inti} has poles at  $\theta=\theta'-\kappa-i\eta_{1}$
and $\theta=\theta'+\kappa+i\eta_{2}$. To calculate the $\kappa$-regularized
part of the integral and evaluate the $\eta_{1,2}\to0$ limit, we
modify the integration contour for $\theta$ to be the sum of the contours
\begin{eqnarray}
\mathcal{C}_{1} & = & \left\{ x-s+i\phi|x\in\left[-\infty,0\right]\right\} \,,\nonumber\\
\mathcal{C}_{2} & = & \left\{ -s+ix|x\in\left[\phi,-\phi\right]\right\} \,,\nonumber\\
\mathcal{C}_{3} & = & \left\{ x-s-i\phi|x\in\left[0,\infty\right]\right\} \,,
\end{eqnarray}
where $s$ and $\phi$ are parameters chosen carefully. We have that $s<\theta'-\kappa$,
$\eta_{1}<\phi$, and $\phi$ has to be smaller than the position
of the branch point in the  function $\hat{K}$. When shifting the contour from the real axis to $\mathcal{C}_3$ we encounter a pole at $\theta=\theta'-\kappa-i\eta_{1}$ and pick up the residue contribution, in a clockwise direction, with the value
\begin{equation}
\label{residue1}
n\int_{0}^{\infty}\frac{\mathrm{d}\theta'}{2\pi}M(\theta',\theta'-\kappa; t)\left[f(2\kappa+i\eta+i\pi)+f(2\kappa+i\eta-i\pi)\right]\,.
\end{equation}
where $\eta:=\eta_1+\eta_2$.
Expanding the integrand in~\eqref{residue1} around $\kappa=-i\eta/2$, sending $\eta\rightarrow 0$, and calling $\theta$ the integration variable, we have 
\begin{equation}
\frac{in}{\kappa+i\frac{\eta}{2}}\int_{0}^{\infty}\frac{\mathrm{d}\theta}{2\pi}\hat{K}^2(\theta)-2n mt\int_{0}^{\infty}\frac{\mathrm{d}\theta}{2\pi} \hat{K}^2(\theta)\sinh\theta-n\int_{0}^{\infty}\frac{\mathrm{d}\theta}{2\pi}\frac{d\hat{K}(\theta)}{d\theta}\hat{K}(\theta)+O(\kappa)\,.
\label{88}
\end{equation}
The first term in \eqref{88} exactly cancels the two-particle form factor singularity, i.e. the first term on the right hand side of (\ref{C22_kappa}).
The second term is remarkably linear in time with coefficient $-\frac{nm\Gamma}{2}$ and $\Gamma$ given in~\eqref{gamma}. The third term vanishes due to $\hat{K}^2(0)=\hat{K}^2(\infty)=0$. We can then finally write
\begin{align}
\label{int_i}
I(t) =&  -\frac{n \Gamma m t }{2}+n\int_{0}^{\infty}\frac{\mathrm{d}\theta'}{2\pi}\int_{\mathcal{C}_{1}\cup\mathcal{C}_{2}\cup\mathcal{C}_{3}}\frac{\mathrm{d}\theta}{2\pi}\frac{M(\theta',\theta';t)H(\theta',\theta)}{2\sinh^2(\frac{\theta-\theta'}{2})}\,,
\end{align}
where the last integral is now well defined as there are no singularities along the contour of integration of the rapidity $\theta$.

It is also possible to extract the large time limit of the integral 
\begin{equation}
\label{intr}
R(t):=n \int_{0}^{\infty}\frac{\mathrm{d}\theta'}{2\pi}\int_{\mathcal{C}_{1}\cup\mathcal{C}_{2}\cup\mathcal{C}_{3}}\frac{\mathrm{d}\theta}{2\pi}\frac{M(\theta',\theta't)H(\theta',\theta)}{2\sinh^2(\frac{\theta-\theta'}{2})}\,,
\end{equation}
which appears in~\eqref{int_i}. The integrand of~\eqref{intr} has a double pole on the real axis of the variable $\theta$, however this can be cured, without spoiling convergence at infinity, by taking a double derivative with respect to time. After taking the double derivative the integration contour for $\theta$ can be lifted back to the real axis. We are then led to consider the large $t$ asymptotics of the  following double integral
\begin{equation}
\label{der_2}
\frac{d^2R(t)}{dt^2}=-4m^2n\int_{0}^{\infty}\frac{d\theta'}{2\pi}\int_{-\infty}^{\infty}\frac{d\theta}{2\pi} \frac{M(\theta',\theta)H(\theta',\theta)(\cosh\theta'-\cosh\theta)^2}{2\sinh^2\left(\frac{\theta-\theta'}{2}\right)}\,.
\end{equation}
This can be done by  standard application of the stationary phase approximation for two-dimensional integrals. There is only one stationary point at $\theta=\theta'=0$; Taylor-expanding the integrand about $\theta=\theta'=0$ gives at leading order
\begin{align}
\frac{d^2 R(t)}{dt^2}=-\frac{n\mu^2}{32 m\pi t^3}+O(t^{-5})\,.
\end{align} 
Integrating back twice we obtain the desired asymptotic for the integral $R(t)$ in~\eqref{int_i} which is 
\begin{equation}
R(t)=-\frac{n\mu^2}{64 m\pi t}+O(t^{-3})\,.
\label{RTRT}
\end{equation}
Notice that when integrating back, we are setting to zero possible terms $O(t)$ and $O(1)$, due to the asymptotic of the original integral. In summary, we have shown that
\beq
I(t)= -\frac{n \Gamma m t }{2}-\frac{n\mu^2}{64 m\pi t}+O(t^{-3})\,.
\eeq
We will revisit this result in subsection \ref{hints2} where we argue that these contributions are nothing but the first non-trivial term in the expansion of the exponential featuring in (\ref{twist_exp}).

\subsubsection{The Contributions $D_{0,4}$ and $D_{4,0}$}
Finally we analyze the contributions $D_{0,4}$ and $D_{4,0}$. Since $D_{4,0}=D_{0,4}^*$, we focus only on  $D_{0,4}$, which is given by  $D_{0,4}=C_{0,4}$ with
\begin{equation}
 \label{C04}
 \tilde{\tau}_n C_{0,4}(t)=-\frac{n}{2}\sum_{j=1}^n\int_{0}^{\infty}\frac{d\theta d\theta'}{(2\pi)^2}N(\theta',\theta; t) F^{11jj}(-\theta',\theta',-\theta,\theta)\,,
\end{equation}
and we defined, analogously to~\eqref{mfun}
\begin{equation}
 \label{nfun}
 N(\theta',\theta; t):=\hat{K}(\theta')\hat{K}(\theta)e^{-2imt(\cosh\theta'+\cosh\theta)}\,.
\end{equation}
The four-particle form factor in~\eqref{C04} can be decomposed as in~\eqref{1stlinethis} by applying  Wick's theorem and the sum over the index $j$ performed by recalling~\eqref{ggg}. By repeating steps similar to those employed in the section~\ref{sec:ksquare}, the integral $C_{0,4}$ can be written as a sum of two terms, namely $C_{0,4}(t):=C_{0,4}^{(1)}(t)+C_{0,4}^{(2)}(t)$. In particular, one obtains
\begin{equation}
 \label{C041}
 C_{0,4}^{(1)}(t)=-\frac{1}{2}\left[ n\int_{0}^{\infty}\frac{d\theta}{2\pi} \hat{K}(\theta)f(-2\theta)e^{-2imt\cosh\theta}\right]^{2}=\frac{1}{2} [C_{0,2}(t)]^2\,,
\end{equation}
and
\begin{equation}
 \label{C042}
 C_{0,4}^{(2)}(t)=\frac{n}{2}\int_{0}^{\infty}\frac{d\theta'}{2\pi}\int_{-\infty}^{\infty}\frac{d\theta}{2\pi}N(\theta',\theta; t)\frac{H(\theta',\theta)}{2\cosh^2\left(\frac{\theta'-\theta}{2}\right)}\,,
\end{equation}
where $H(x,y)$ is the same function given in~\eqref{funH}. By applying the stationary phase approximation we can estimate the large time limit of ~\eqref{C042}.  This gives another $O(t^{-3})$ contribution (since (\ref{C041}) is also of $O(t^{-3})$), namely
\begin{equation}
 C_{0,4}^{(2)}(t)+C_{4,0}^{(2)}(t)=\frac{b_n\mu^2}{32\pi}\frac{\sin(4mt)}{(mt)^3}+O(t^{-7/2})\,,
\end{equation}
with
\begin{equation}
b_n=\frac{2+n^2-12\cot\left(\frac{\pi}{n}\right)\csc\left(\frac{\pi}{n}\right)}{48n}\,.
\end{equation}
This closes our calculation of the branch point twist field one-point function at $O(K^2)$.

\subsubsection{The Complete Formula at $O(K^2)$ for the Twist Field One-point Function}

We can finally summarize the result for the twist field one-point function up to $O(K^2)$
\beqa
\label{final_twist2}
\frac{{}_n\bra \Omega |\TT(0,t)|\Omega\ket_n}{{}_n\bra \Omega|\Omega\ket_n}&=&\tilde{\tau_n}\left[1+A+C_{2,0}(t)+C_{0,2}(t)-\frac{ n\Gamma mt}{2}+\right. \nonumber \\
&&\qquad\left.+\frac{(C_{2,0}(t)+C_{0,2}(t))^2}{2}+R(t)+ C_{4,0}^{(2)}(t)+C_{0,4}^{(2)}(t)\right]+O(K^3)\,,
\eeqa
where $A$ is given in~\eqref{A_const}, $C_{2,0}(t)+C_{0,2}(t)$ in~\eqref{C2002}, $\Gamma$ in \eqref{gamma},  $R(t)$ in~\eqref{intr} and $C_{0,4}^{(2)}(t)=[C_{4,0}^{(2)}(t)]^*$ in \eqref{C042}.  Note also that 
\beq
\frac{(C_{2,0}(t)+C_{0,2}(t))^2}{2}=C_{0,4}^{(1)}(t)+C_{4,0}^{(1)}(t)+I'(t)\,,
\eeq
 with  $I'(t)$ given in (\ref{Iprime}) and the other terms in (\ref{C041}). The presence of the contributions $1+C_{2,0}(t)+C_{0,2}(t)
+\frac{(C_{2,0}(t)+C_{0,2}(t))^2}{2}$ suggests that these and higher-order terms may arise from the exponentiation of $C_{2,0}(t)+C_{0,2}(t)$. 
Indeed, it is possible to argue that all terms in the expansion (\ref{final_twist2}) exponentiate. 
We will give a simple argument towards this conclusion in subsection~\ref{hints2}.

Reinserting the cut-off dependence $\varepsilon^{\Delta_n}$ in~\eqref{final_twist2}, taking the logarithm, expanding its argument up to $O(K^2)$ and dividing by $1-n$ we obtain the following expression up to $O(K^2)$ for the R\'enyi entropies
\begin{equation}
\label{final_Renyi2}
S_n(t)-{\tilde{S}_{0,n}}=\frac{1}{1-n}\left[A+C_{2,0}(t)+C_{0,2}(t)-\frac{ n\Gamma mt}{2}+R(t)+C_{0,4}^{(2)}(t)+C_{0,4}^{(2)}(t)\right]+O(K^3)\,,
\end{equation}
where $\tilde{S}_{0,n}$ is the R\'enyi entropy in the ground state of the post-quench Hamiltonian i.e.
\begin{equation}
\label{pqVEV}
\tilde{S}_{0,n}=\frac{\log(\varepsilon^{\Delta_n}\tilde{\tau}_n)}{1-n}\,. 
\end{equation}
Expanding~\eqref{final_Renyi2} for $mt\gg1$, we obtain the result quoted in~\eqref{mainres}.

\subsection{An Argument Towards Exponentiation at Higher Orders}
\label{hints2}
A simple combinatorial argument can be provided to show that all the terms in the expansion (\ref{final_twist2}) exponentiate. 
In other words, they result from the expansion of an exponential at order $O(K^2)$. The exponent will receive $O(K^3)$ and higher corrections which we will not investigate in this paper. Note that in \cite{SE} an argument was given for the exponentiation of the term $-\Gamma m t$ (the equivalent of our $-\frac{\Gamma n mt}{2}$ term but for the order parameter). An entirely similar argument can be given for the branch point twist field to show the exponentiation of this term. However, we find that exponentiation is a much more general feature of the one-point function, extending to other terms at $O(K^2)$ as well. 
As we will see, our calculation does not use any special properties of the branch point twist field form factors, apart from their Pfaffian structure. Therefore we expect the same exponentiation to occur for the order parameter $\frac{\bra \Omega | \sigma (0,t)| \Omega\ket}{ \bra \Omega | \Omega\ket}$. 

Examining the generalization of the crossing relation \eqref{bringover} and the Wick contraction nature of the form factor expressions \eqref{paf} and \eqref{K}, it is natural to expand the $C_{2k,2l}(t)$ functions as sums of products of connected contributions, that is
\beq
C_{2k,2l}(t) =  \sum_{ \{n_{i,j}\}} \prod_{i,j=0}^\infty \frac{\left(C_{2i,2j}^{c}(t)\right)^{n_{i,j}}}{n_{i,j}!}\,, \label{eq:connected_expansion}
\eeq
where $C_{2i,2j}^c(t)$ are related to integrals of ``connected" matrix elements, which are defined recursively from the condition of not being factorizable into other connected expressions. The $n_{i,j}$ are non-negative integers that  satisfy the constraints $\sum_{i,j=0}^\infty i \, n_{i,j} =k$ and $\sum_{i,j=0}^\infty j \,n_{i,j} =l$. By inverting the expansion \eqref{eq:connected_expansion}, for the first few connected terms we get for instance
\beqa
C_{2,0}^c(t)=C_{2,0}(t) \,, &\quad & C_{0,2}^c(t)=C_{0,2}(t) \,, \\ 
C_{4,0}^c(t)=C_{4,0}(t)- \frac{1}{2}\left(C_{2,0}(t)\right)^2 \,, &\quad & C_{0,4}^c(t)=C_{0,4}(t)- \frac{1}{2}\left(C_{0,2}(t)\right)^2 \,, \\
C_{2,2}^c(t) =  C_{2,2}(t)- C_{2,0}(t) C_{0,2}(t)\,. 
\eeqa
These new combinations of terms are immediately recognizable from our earlier computation. For instance, 
$C_{0,4}^c(t)$ is nothing but $C_{0,4}^{(2)}(t)$ defined in \eqref{C042}, and $C_{2,2}^c(t)$ is obtained from $C_{2,2}(t)$ after subtracting the term $I'(t)$ defined in \eqref{Iprime}. 
The norm of the initial state $|\Omega\ket_n$ admits an analogous expansion 
\beq
Z_{2k} =  \sum_{ \{\tilde{n}_{i}\}} \prod_{i=0}^\infty \frac{\left(Z_{2i}^{c}\right)^{\tilde{n}_{i}}}{\tilde{n}_{i}!}\,, 
\eeq
where $\sum_{i=0}^\infty i\, \tilde{n}_i=k$. To calculate the regular terms of the one-point function $D_{2k,2l}(t)$, see eq. \eqref{eq:D_2k_2l_original}, we need to calculate the inverse of the norm defined by the condition $\sum_{km,p=0}^\infty Z_{2k} \tilde{Z}_{2p}=1$. From observation of the first few terms of the inverse of the norm, we expect its connected expansion to have the form
\beq
\tilde{Z}_{2k} =  \sum_{ \{\tilde{n}_{i}\}} \prod_{i=0}^\infty \frac{\left(-Z_{2i}^{c}\right)^{\tilde{n}_{i}}}{\tilde{n}_{i}!}\,. \label{eq:inverse_norm_connected}
\eeq
In the following, we are only focusing on terms of the one-point function, that contain connected matrix elements of at most  $O(K^2)$, i.e. we consider only terms where powers $n_{i,j}=0$ for $i+j> 2$ and $\tilde{n}_i=0$ for $i>1$. With this assumption \eqref{eq:connected_expansion} takes the form 
\beq
C_{2(k+l), 2k}=\sum_{r=0}^{k}\frac{\left(C_{2,2}^c(t)\right)^{k-r}}{(k-r)!} L^{r+l}(t) R^{r}(t) \,,
\eeq
where 
\beq
L^{k}(t)   =  \sum_{p=0}^{\lfloor\frac{k}{2} \rfloor} \frac{\left(C_{4,0}^c(t)\right)^{p}}{p!}\frac{\left(C_{2,0}^c(t)\right)^{k-2p}}{(k-2p)!}\,,  \qquad \mathrm{and} \qquad
R^{k}(t)   =  \sum_{q=0}^{\lfloor\frac{k}{2} \rfloor} \frac{\left(C_{0,4}^c(t)\right)^{q}}{q!}\frac{\left(C_{0,2}^c(t)\right)^{k-2q}}{(k-2q)!} \,,    
\eeq
with $\lfloor . \rfloor$ denoting the integer part. Plugging these formulas and \eqref{eq:inverse_norm_connected} into $D_{2(k+l),2k}$ using the form \eqref{eq:D_2k_2l_original}, and exchanging the order of the summations leads to 
\beq 
D_{2(k+l),2k}(t)=\sum_{r=0}^{k} \frac{ \left(D_{2,2}^c(t)\right)^{k-r}}{(k-r)!}L^{r+l}(t)R^r(t)\,, 
\eeq 
where the combination $D_{2,2}^c (t)=C_{2,2}^c (t) -Z_2^c$ is both connected and regular. 
Similar results hold for $D_{2k,2(k+l)}(t)$ and $D_{2k,2k}(t)$, hence the one-point function \eqref{d_series} takes the form  
\beqa
\!\!\!\frac{{}_n \bra \Omega| \TT(0,t)|\Omega\ket_n}{{}_n \bra \Omega|\Omega\ket_n}&=&\tilde{\tau}_n \left\{ \sum_{k=0}^{\infty} \sum_{r=0}^{k} \frac{ \left(D_{2,2}^c(t)\right)^{k-r}}{(k-r)!} \left[\sum_{l=1}^{\infty} \left[L^{r+l}(t)R^r(t)+L^{r}(t)R^{r+l}(t)\right] +L^r(t) R^r(t)\right]\right.\nonumber\\
&& + \left. O(K^3)\right\} \,.
\eeqa
Further manipulation of the order and range of the summations, allows us to write the one-point function as
\beqa
\frac{{}_n \bra \Omega| \TT(0,t)|\Omega\ket_n}{{}_n \bra \Omega|\Omega\ket_n} &=&\tilde{\tau}_n  \left\{e^{D_{2,2}^c(t)}\sum_{r=0}^{\infty} L^r(t)\sum_{s=0}^{\infty} R^s(t)+ O(K^3)\right\} \nonumber \\
&=&\tilde{\tau}_n  \, e^{D_{2,2}^c(t)+D_{2,0}^c(t)+D_{4,0}^c(t)+D_{0,2}^c(t)+D_{0,4}^c(t)+ O(K^3)}  \,,
\eeqa
where $D_{2i,0}^c (t)=C_{2i,0}^c (t)$, $D_{0,2i}^c (t)=C_{0,2i}^c (t)$, since these terms are regular without any subtraction.  
Note that the terms in the exponent are precisely those inside the bracket in (\ref{final_Renyi2}).
With this we showed, using the assumption \eqref{eq:inverse_norm_connected}, that the one-point function exponentiates up to $O(K^2)$ terms. The expression is regular, since all the singularities are cancelled as explained in detail in the previous sections. 

Given the simplicity of the Ising form factors, 
we expect exponentiation to occur also at higher orders in $K$, and we are planning to investigate this further in the future.  

\section{Perturbation Theory in the Quench Parameter}
\label{pert_theory}
Integrable model perturbation theory was developed in \cite{PIQFT} for  the study of integrable models subject to a small integrability-breaking perturbation. 
The case considered there was translation invariant in time.
In a non-equilibrium protocol, such as a quench, the field theory action is no longer time-translation invariant. In \cite{PQ1} it was then observed that requiring factorization of the scattering \textit{at all times} for such an action is consistent only if the latter is free.  An approach to tackle the quench problem was also proposed in which the state in the Heisenberg picture after the quench could be expanded perturbatively in the quench parameter over the pre-quench quasi-particle basis. The approach requires the pre-quench theory to be integrable but allows for considering integrability breaking protocols. 

 Let us first review the main results of \cite{PQ1}. Consider an integrable quantum field theory with ground state $|0\ket$ and action $\mathcal{A}_0$.  At time $t=0$ the system is  quenched and  from $t=0$ onwards it is described by the new action 
\beq
\label{action}
\mathcal{A}=\mathcal{A}_0-\lambda \int_0^\infty dt \int_{-\infty}^\infty dx \, \Psi(x,t)\,,
\eeq
where $\Psi(x,t)$ is some local field. In the interaction picture, with  respect to the Hamiltonian of the pre-quench theory, the state of the system at infinite time after the quench is the time ordered exponential
\begin{equation}
\label{state_p}
|\psi_0\ket= \lim_{t\rightarrow\infty} T \left[\exp\left(-i \lambda \int_0^t ds \int_{-\infty}^\infty dx \, \Psi(x,s) \right)\right]|0\ket\,. 
\end{equation}
The state $|\psi_0\rangle$ in~\eqref{state_p}  can then be expanded perturbatively in $\lambda$ over the basis of the out-states of the pre-quench theory. $k$-particle states of this type  are denoted by  $|\theta_1,\ldots, \theta_k\ket_{\text{out}}$,  with $\theta_1<\theta_2 < \cdots <\theta_k$, being the rapidities. It can then be shown that integrability of the pre-quench theory allows for relaxing the constraint of ordering on the rapidities in the expansion over the out-states. In fact, the expansion~\cite{PQ1}   
\beqa
|\psi_0\rangle=|0\ket + \lambda \sum_{k=1}^\infty  \frac{2\pi}{k!} \int_{-\infty}^\infty  \prod_{i=1}^k \frac{d \theta_{i}}{2\pi}~\frac{ \delta(\sum_i P_0(\theta_i)) [F_k^{\Psi}(\theta_1,\ldots,\theta_k)]^*}{\sum_{i} E_0(\theta_i)} |\theta_1\ldots \theta_k\ket+O(\lambda^2)\,, 
\label{psi0}
\eeqa
represents the state in the pre-quench basis in the Heisenberg picture at all times after the quench, up to first order in $\lambda$. $E_0(\theta)=m_0 \cosh \theta$ and $P_0(\theta)=m_0 \sinh \theta$ are the pre-quench energy and momenta of the particles,
 and
\beq
F_k^{\Psi}(\theta_1,\ldots,\theta_k):=\bra 0| \Psi(0,0)|\theta_1,\ldots,\theta_k\ket\,,
\eeq
is a $k$-particle form factor of the local field $\Psi$, calculated in the pre-quench quasi-particle basis. This state can then be employed  to compute perturbative corrections to the one-point function of any local field $\Phi$ after the quench. These are found to be 
\beqa
\delta \bra \Phi (t)\ket& =& \bra \psi_0|\Phi(0,t)|\psi_0 \ket-\bra 0|\Phi(0,0)|0\ket =\lambda \sum_{k=1}^\infty  \frac{2\pi}{k!}\nonumber \\
&&\times   \int_{-\infty}^\infty \prod_{i=1}^k \frac{d \theta_{i}}{2\pi}~\frac{ \delta(\sum_i P_0(\theta_i)) }{\sum_{i} E_0(\theta_i)}2\mathrm{Re}\left[[F_k^{\Psi}(\theta_1,\ldots,\theta_k)]^* F_k^{\Phi}(\theta_1,\ldots,\theta_k) e^{-i \sum_{i=1}^k E_0(\theta_i) t}\right] \nonumber\\
&& +C_\Phi+O(\lambda^2)\,, 
\label{delta}
\eeqa
where~\cite{PQ2}
\beqa
C_\Phi=- \lambda \sum_{k=1}^\infty   \frac{2\pi}{k!}  \int_{-\infty}^\infty \prod_{i=1}^k \frac{d \theta_{i}}{2\pi}~\frac{ \delta(\sum_i P_0(\theta_i)) }{\sum_{i} E_0(\theta_i)}  2 \mathrm{Re}\left[[F_k^{\Psi}(\theta_1,\ldots,\theta_k)]^* F_k^{\Phi}(\theta_1,\ldots,\theta_k) \right]\,,
\label{cphi}
\eeqa
is a constant which  is introduced  to ensure that $\delta \bra \Phi(0) \ket=0 $ at first order in perturbation theory.

\subsection{Perturbation Theory for the Entanglement Entropy}
In order to calculate the quantity
 $\delta \bra \TT(t) \ket_n$, defined similarly to (\ref{delta}), we shall work in a replica version of~\eqref{action}; however this introduces a few changes. As discussed in section \ref{sec:tech} particles are  labelled by a replica index $j=1,\ldots,n$ and we will denote  the replicated normalized pre-quench ground state   by $|0\ket_n=\otimes_n |0\ket$. By repeating the steps leading to~\eqref{psi0}, it follows that the first order expansion of the state of the system in the replica theory after the quench is
\beq
|\psi_0\ket_n= |0\ket_n + \lambda n \sum_{k=0}^\infty \frac{2\pi}{k!} \int_{-\infty}^\infty \prod_{i=1}^k \frac{d \theta_{i}}{2\pi}~\frac{ \delta(\sum_i P_0(\theta_i)) [F_k^{\Psi}(\theta_1,\ldots,\theta_k)]^*}{\sum_{i} E_0(\theta_i)} |\theta_1\ldots \theta_k\ket_{1\ldots,1;n}+O(\lambda^2)\,.
\label{replica_state}
\eeq
The expression \eqref{replica_state} is essentially identical to (\ref{psi0}) except for the prefactor $n$,  which takes into account the sum over the replicas. Such a sum is however trivial since the local operator $\Psi$ when insterted in the $j$-th replica  has only non-vanishing form factors among particles with copy index $j$.  Similarly, the generalization of (\ref{delta}) and (\ref{cphi}) for the twist field is also  straightforward and given by
\beqa
\delta \bra \TT (t)\ket_n &=&  {}_n\bra \psi_0|\TT(0,t)|\psi_0 \ket_n- {}_n \bra 0|\TT(0,0)|0\ket_n = \lambda n \sum_{k=1}^\infty \frac{2\pi}{k!}  \nonumber  \\
&&\times  \int_{-\infty}^\infty \prod_{i=1}^k \frac{d \theta_{i}}{2\pi}~\frac{ \delta(\sum_i P_0(\theta_i)) }{\sum_{i} E_0(\theta_i)}2\mathrm{Re}\left[[F_k^{\Psi}(\theta_1,\ldots,\theta_k)]^* F_k^{1\ldots 1}(\theta_1,\ldots,\theta_k) e^{-i \sum_{i=1}^k E_0(\theta_i) t}\right]\nonumber \\
&&+C_\TT^n+O(\lambda^2)\,,
\label{delta2}
\eeqa
where
\beqa
C_\TT^n=- \lambda n \sum_{k=1}^\infty  \frac{2\pi}{k!}  \int_{-\infty}^\infty \prod_{i=1}^k \frac{d \theta_{i}}{2\pi}~\frac{ \delta(\sum_i P_0(\theta_i)) }{\sum_{i} E_0(\theta_i)}  2 \mathrm{Re}\left[[F_k^{\Psi}(\theta_1,\ldots,\theta_k)]^* F_k^{1\ldots 1}(\theta_1,\ldots,\theta_k) \right]\,,
\label{cphi2}
\eeqa
and $F_k$ are the form factors defined in~(\ref{paf}), see section~\ref{sec:tech}. The entanglement entropy may then be computed at first order in perturbation theory as
\beq
S_n(t)=\frac{\log[\varepsilon^{\Delta_n}(\tau_n + \delta \bra \TT (t)\ket_n)]}{1-n}=\frac{\log(\varepsilon^{\Delta_n}\tau_n)}{1-n} + \frac{\delta \bra \TT (t)\ket_n}{\tau_n(1-n)}+O(\lambda^2)\,.
\label{EEdy2}
\eeq
Finally, we can define the first order correction to the R\'enyi entropies as 
\begin{equation}
\label{first_Reny}
\delta S^{1}_n(t):=\frac{\delta \bra \TT (t)\ket_n}{\tau_n(1-n)}\,.
\end{equation}
Notice that in the perturbative approach,  the pre-quench VEV (i.e. $\tau_n$) appears at the denominator of \eqref{first_Reny}. 

\subsection{Entanglement Entropy Oscillations after a Small Mass Quench}
Let us now evaluate (\ref{first_Reny}) for a mass quench in the Ising field theory. In this case the field $\Psi(x,t)$ is the energy field, denoted by $\varepsilon(x,t)$, which has only a non-vanishing two-particle form factor with the pre-quench basis. Indeed, the pre-quench action $\mathcal{A}_0$ in \eqref{action} is obtained by perturbing the conformal invariant UV fixed point by the energy operator itself. The two-particle form factor, suitably normalized reads 
\beq
\label{eff}
F^{\varepsilon}_2(\theta)=-2m_0 i \sinh\frac{\theta}{2}\,.
\eeq
 With the normalization choice for the energy form factor given in~\eqref{eff}, we can directly identify~\cite{PQ2} $\lambda$ in~\eqref{delta2} with $\delta m\ll 1$, given in~\eqref{mu}. From~\eqref{delta2}--(\ref{EEdy2}), and also recalling~\eqref{full}, the first order correction in $\delta m$ to the R\'enyi entropies after the quench can be easily calculated:
\begin{equation}
 \delta S^{1}_n(t)=\frac{1}{1-n} \frac{\delta m}{m_0}  \int_{-\infty}^\infty 
\frac{d\theta}{4 \pi \cosh^2 \theta} \frac{ \sinh\theta \sinh \frac{\theta}{n} 
\cos\frac{\pi}{2n}}{\sinh\frac{i\pi-2\theta}{2n} \sinh\frac{i\pi+2\theta}{2n}}\cos(2   m_0 t \cosh \theta)+\frac{C^n_{\mathcal{T}}}{\tau_n(1-n)}\,.
\label{21}
\end{equation} 
The constant $C_{\mathcal{T}}^n$ can be  determined exactly in this case~\cite{PQ2} and it turns out to be
 \begin{equation}
  \label{Cphi_exact}
  \frac{C_{\mathcal{T}}^n}{\tau_n}=\frac{\delta m}{m_0} \Delta_n \,,
 \end{equation}
where $\Delta_n$ is the scaling dimension of the twist field in~\eqref{stwist}. Observing that at first order in the quench parameter
\begin{equation}
\tilde{\tau}_n=\tau_n\left(1+\Delta_n\frac{\delta m}{m_0}\right)+O\left(\frac{\delta m^2}{m_0^2}\right),
 \end{equation}
 then from~\eqref{EEdy2} and~\eqref{21},  the large time limit at first order in perturbation theory for the R\'enyi entropies finally follows
 \begin{equation}
 \label{reny_pert}
  S_n(0)+\delta S^1_n(t)=\frac{\log(\varepsilon^{\Delta_n}\tilde{\tau}_n)}{1-n}+\frac{1}{8 \sqrt{\pi} n(1-n)} \frac{\delta m}{m_0}  \frac{
\cos\frac{\pi}{2n}}{\sin^2\frac{\pi}{2n}} \frac{\cos(2m_0t-\frac{\pi}{4})}{(m_0t)^{\frac{3}{2}}}+O(t^{-5/2})\,.
 \end{equation}
Eq.~\eqref{reny_pert} reproduces the main result in~\eqref{mainres}, up to $O(\mu)$  as anticipated in section~\ref{sec:sum}. By expanding $\hat{K}$ around $\mu=0$, it is actually easy to verify that eq.~\eqref{21} coincides with the first order of~\eqref{final_Renyi2} at all times.

\section{Lattice Results and Numerical Study in the Scaling Limit}
\label{sec:num}
In this section we present a detailed comparison of the field theory results obtained in section~\ref{massquench} against lattice numerical calculations in the Ising spin chain with Hamiltonian
\begin{equation}
\label{ising2}
 H_{\text{Ising}}(h)=-J\sum_{i=1}^N(\sigma_i^x\sigma_{i+1}^x+h\sigma_{i}^z)\,.
\end{equation}
In the following, as already anticipated in the introduction, the lattice spacing will be denoted by $a$. In a lattice model, it is not possible to access directly the R\'enyi entropies $S_{n}(t)$ for a semi-infinite interval after a quench. Numerical techniques, based on the correlation matrix, are however known for calculating the R\'enyi entropies $S_{n}^{L}(t)$ of a subsystem of $L$ neighbouring sites with physical length $\ell:=La$, embedded into an infinite system (i.e. in the limit $N\rightarrow\infty$ in~\eqref{ising2}). To extract the semi-infinite R\'enyi entropies, which we determined analytically in section~\ref{massquench}, we then assume the validity of the same clustering property that holds  for the spin-operator two-point function. The clustering property translates  for the R\'enyi entropies into 
\begin{equation}
 \label{clustering_reny}
 \lim_{L\rightarrow\infty}S_{n}^{L}(t)=2S_{n}(t)\,.
\end{equation}
As also discussed in section~\ref{sec:intro}, from a field theoretical perspective (see for instance \eqref{cluster}) eq.~\eqref{clustering_reny} is a consequence of locality of the branch point twist field, nevertheless it constitutes  a  non-trivial and new prediction when applied to the lattice model.

\subsection{Correlation Matrix}
\label{sec:correlationmatrix}
For the Ising spin chain, the time evolution of correlation functions and entropies can be calculated using the restricted correlation matrix of a subsystem $A$ of $L$ sites~\cite{vidal03, EEquench}, embedded into an infinite system
\begin{equation}
    \label{eq:Cmatr}
    \Gamma_L^A=\left[\begin{array}{cccc}
        \Pi_0 & \Pi_{-1} & \cdots & \Pi_{1-L} \\
        \Pi_{1} & \Pi_{0} &  & \vdots  \\
        \vdots & & \ddots & \vdots \\
        \Pi_{L-1} & \cdots & \cdots & \Pi_0
    \end{array}\right]\,,\qquad \mathrm{with} \qquad \Pi_j=\left[\begin{array}{cc}
        -f_j & g_j \\
        -g_{-j} & f_j
    \end{array}\right]\,,
\end{equation}
where\footnote{Note that the matrix $\Gamma_L^A$ has nothing to do with the decay rate $\Gamma$ introduced earlier in (\ref{gamma}). Both these notations have been previously used in the literature so we maintain them here.}
\begin{equation}
    \label{eq:fg_def}
    \begin{split}
        g_j & =  \frac{1}{2\pi}\int_{-\pi}^{\pi}\mathrm{d}\varphi \mathrm{e}^{-\mathrm{i}\varphi j}\mathrm{e}^{-\mathrm{i}\theta_{\varphi}}\left(\cos \Phi_{\varphi}-\mathrm{i}\sin \Phi_{\varphi}\cos 2\epsilon_{\varphi}t\right)\,, \\
        f_j & =  \frac{\mathrm{i}}{2\pi}\int_{-\pi}^{\pi}\mathrm{d}\varphi \mathrm{e}^{-\mathrm{i}\varphi j}\sin \Phi_{\varphi}\sin 2\epsilon_{\varphi}t\,,
    \end{split}
\end{equation}
and, for the Ising chain, we have 
\begin{equation}
    \label{eq:bog_defs}
    \begin{split}
        \epsilon_{\varphi} & =\frac{1}{a}\sqrt{(1+a m-\cos \varphi)^2+\sin^2 \varphi}\,, \\
        \epsilon^0_{\varphi} & =\frac{1}{a}\sqrt{(1+a m_0-\cos \varphi)^2+\sin^2 \varphi}\,, \\
        \mathrm{e}^{-\mathrm{i}\theta_{\varphi}} & = \frac{\cos \varphi -(1+a m)-\mathrm{i}\sin \varphi}{a\epsilon_{\varphi}}\,, \\
        \sin \Phi_{\varphi} & = \frac{\sin \varphi(a m_0-a m)}{a^2\epsilon_{\varphi}\epsilon^0_{\varphi}}\,, \\
        \cos \Phi_{\varphi} & = \frac{1-\cos \varphi(a m_0+a m+2)+(1+a m)(1+a m_0)}{a^2\epsilon_{\varphi}\epsilon^0_{\varphi}}\,.
    \end{split}
\end{equation}
Here, we already rewritten the transverse fields $h_0$ and $h$ in terms of the pre- and post-quench masses  defined in the scaling field theory by:  $h_0=1+a m_0$, $h=1+am$. Since $m_0,m$ are positive, the lattice calculations will be performed in the paramagnetic phase, the results should hold also in the ferromagnetic phase by duality. We also set the speed of light to $v=1$, therefore, according to~\eqref{scalinglimit}, $J=\frac{1}{2a}$. 

The matrix $\Gamma_L^A$ has $2L$ purely imaginary eigenvalues $\pm\mathrm{i}\nu_k,\,k=1,\dots,L$, and the $2^{L}$ eigenvalues of the reduced density matrix $\rho_A$
matrix have the form
\begin{equation}
\lambda_j=\frac{1}{2^{L}}\prod_{k=1}^{L}\left(1+(-1)^{a^{(j)}_k}\nu_{k}\right)\,,
\end{equation}
where $a^{(j)}_k\in \{0,1\}$. A straightforward calculation gives the R\'enyi entropies for an interval of length $L$
\begin{equation}
    \label{eq:renyi_pol}
    S^{L}_n(t)=\frac{1}{1-n}\log \mathrm{Tr}\rho_A^n =\frac{1}{2(1-n)}\mathrm{Tr}\log\left(P_n(\mathrm{i}\Gamma_L^A) \right)\,,
\end{equation}
where the polynomials are
\begin{equation}
    \label{eq:rpoly}
    P_n(x)=\left(\frac{1+x}{2}\right)^n+\left(\frac{1-x}{2}\right)^n\,.
\end{equation}
Therefore the R\'enyi entropies can be easily calculated numerically by diagonalizing the correlation matrix.

\subsection{Linear Growth in the Scaling Limit}

Exact lattice results are available~\cite{FC} for  the leading, linear in time, contribution  to the R\'enyi entropies $S_{n}^{L}(t)$. The linear growth is obtained in the regime $1\ll t\ll L$, while for $t\gg L$, according to a semi-classical quasi-particle picture the R\'enyi entropies saturate to a value proportional to the size $L$ of the interval.

We will then compare the field theoretical results of section~\ref{massquench} valid up to the second order in the quench parameter $\mu$, with the scaling limit of the lattice predictions in~\cite{FC}.
Computationally, see again~\eqref{scalinglimit} and the remarks below~\eqref{eq:bog_defs}, the scaling limit is defined as follows: replace lattice quantities according to~\eqref{eq:bog_defs}, then  introduce the continuum momentum variable $p$ substituting  $\varphi:=p a$, and $0\leq p \leq \frac{2\pi}{a}$,  and eventually take the limit $a\rightarrow 0$. The  mathematical operation will be denoted by  the shorthand notation $\lim_{\text{scal}}$. For instance, it is easy to verify that 
\beq
\label{rel_dis}
\lim_{\text{scal}}\epsilon_{\varphi}:=E_m(p)=\sqrt{m^2+p^2}\,.
\eeq 
The main result in~\cite{FC}, evaluated for $L\rightarrow\infty$ then reads
\beq
 \lim_{L\rightarrow\infty}S^{L}_{n,lin}(t)=\frac{{2}t}{1-n}\int_{0}^\pi \frac{d\phi}{\pi} |\epsilon'_\varphi| \log(P_n(\cos\Phi_{\varphi}))\,,
\label{chaindel}
\eeq
 with $\epsilon'_\varphi:=\frac{d \epsilon_\varphi}{d \varphi}$ and $P_n$ as in~\eqref{eq:rpoly}. The $lin$ subscript indicates that the formula only captures the linear growth part of the entanglement. With the definition
 \begin{equation}
  \label{def_cos_sl}
  \zeta(p):=\frac{mm_0+p^2}{E_{m_0}(p)E_{m}(p)}\,,
 \end{equation}
the scaling limit of~\eqref{chaindel} is thus
\beq
\lim_{\text{scal}}\lim_{L\rightarrow\infty}S_{n,lin}^{L}(t)= \frac{2t}{\pi(1-n)}\int_0^\infty \frac{dp~p}{E_m(p)}\log  \left[\left(\frac{1+\zeta(p)}{2}\right)^n+ \left(\frac{1-\zeta(p)}{2}\right)^n\right].
\label{scalingEE}
\eeq
Expanding~\eqref{scalingEE} for small quenches (i.e. $m=m_0+\delta m$) and substituting $p=m_0\sinh\theta$, which is consistent at second order in $\delta m$, we find
\beq
\label{scal_renyi}
 \lim_{\text{scal}}\lim_{L\rightarrow\infty}S^{L}_{n,lin}(t)= \frac{n t\, \delta m^2}{2\pi m_0(n-1)}\int_0^\infty d\theta\, \frac{\tanh^3 \theta}{\cosh \theta}+O(\delta m^3)= \frac{n m_0 t\mu^2}{3\pi(n-1)}+O(\mu^3)\,,
\eeq
where we used the definition~\eqref{mu}. It then follows, as expected according to~\eqref{clustering_reny}, that the result in~\eqref{scal_renyi}  is precisely {\it twice} the leading large-time asympotics obtained expanding \eqref{mainres}  for a small quench, see in particular \eqref{expgamma}. By considering the limit $n\rightarrow 1$ in~\eqref{chaindel}, the scaling limit of the von Neumann entropy turns out to be
\begin{equation}
\label{scalingEEint}
 \lim_{\text{scal}}\lim_{L\rightarrow\infty}S_{1,lin}^{L}(t)=-\frac{2t}{\pi}\int_{0}^{\infty}\frac{dp~p}{E_m(p)}\left[\left(\frac{1+\zeta(p)}{2}\right)\log\left(\frac{1+\zeta(p)}{2}\right)+\left(\frac{1-\zeta(p)}{2}\right)\log\left(\frac{1-\zeta(p)}{2}\right)\right]\,,
\end{equation}
and expanding for small quenches
\begin{equation}
\label{scalingEE2}
\lim_{\text{scal}}\lim_{L\rightarrow\infty}S_{1,lin}^{L}(t)=  -\frac{t \,\delta m^2 \log\left(\frac{\delta m^2}{m_0^2}\right)}{3\pi m_0}+O(\delta m^2)\,.
\end{equation}
In the scaling limit, and for small quenches, the  von Neumann entropy determined in~\cite{FC} is dominated by a term $O(\delta m^2 \log \delta m)$   and therefore is not analytic in the quench parameter. This  unexpected result provides another indication that the limit $n\rightarrow 1$ in the R\'enyi entropies does not commute with a pertubative expansion in $\delta m$. Incidentally, the emergence of logarithmic corrections to the expectation values of certain fields in the massive Ising field theory is compatible with previous studies,  such as~ \cite{Levi}. It is generally due to ambiguities in the definition of some local  operators: For instance in the Ising field theory, the energy field  can be regarded as a linear combination of the usual fermion bilinear and a term proportional to the identity field with proportionality constant equal to the mass.

\subsection{Numerical Evaluation of the Correlation Matrix}
\label{sec:numerics}
In this section we test the field theory results against the numerical results on the lattice, where one can directly diagonalize the correlation matrix~\eqref{eq:Cmatr} and calculate the entropies as shown in section~\ref{sec:correlationmatrix}.

We expect that the results match in the scaling limit defined in the previous section and for small quenches. Therefore we chose the transverse field to be close to the critical value i.e. $m, m_0 \ll 1$ and the quench $\delta m \ll 1$. Then the scaling limit can be carried out by decreasing $a$ while keeping the physical subsystem size $\ell=La$ fixed. Then one can extrapolate to $a=0$  taking into account corrections to the scaling of the entropies as discussed in Appendix~\ref{app:a}.

This method gives the entanglement entropies of a finite subsystem, which is proportional to the logarithm of the two point function of branch point twist fields. Therefore one needs to consider large enough subsystem sizes in order to observe clustering, namely factorization into a one-point function squared. All our numerical results show excellent agreement with analytical predictions up to a factor two due to clustering. We note that for larger subsystems and times the evaluation of~\eqref{eq:Cmatr} gets harder due to the highly oscillatory integrands in~\eqref{eq:fg_def}. 

As recalled in~\eqref{scal_eq}, in a massive field theory  the logarithmic divergence of the von Neumann entropy is encoded in the term
\beq
S_1 = -\frac{c}{6}\log{m a}+O(1)\,,
\eeq
where $c$ is the central charge of the ultraviolet CFT and the $O(1)$ corrections are discussed in Appendix~\ref{app:a}. The scaling limit technique can be then used to extract the central charge $c$. For instance at values of the mass $m=0.04$ we obtained $c=0.50195(3)$, which is  very close to the theoretical value $c=\frac{1}{2}$. The central charge extrapolation provides a means to numerically probe the scaling regime of the Ising spin chain. We found that differences of entropies calculated at different times do not have any divergences in the scaling limit as expected. More details on the numerical results corresponding to the scaling limit can be found in Appendix~\ref{app:a}.

\subsubsection{Saturation and  Oscillations}

\begin{figure}[t]
    \centering
    \includegraphics[width=7.5cm]{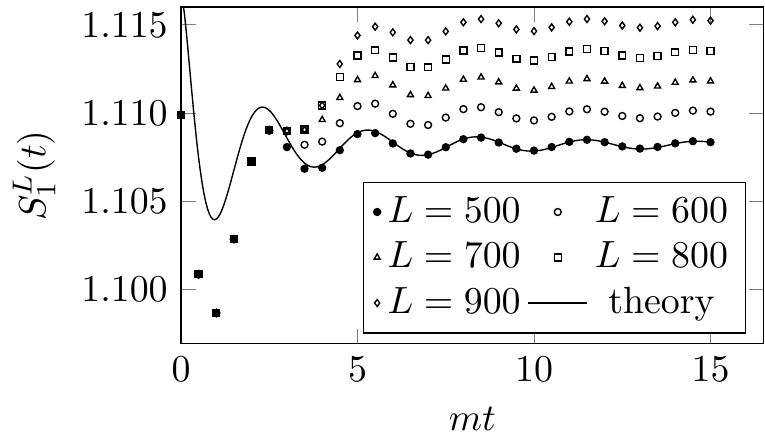} 
    \includegraphics[width=7.5cm]{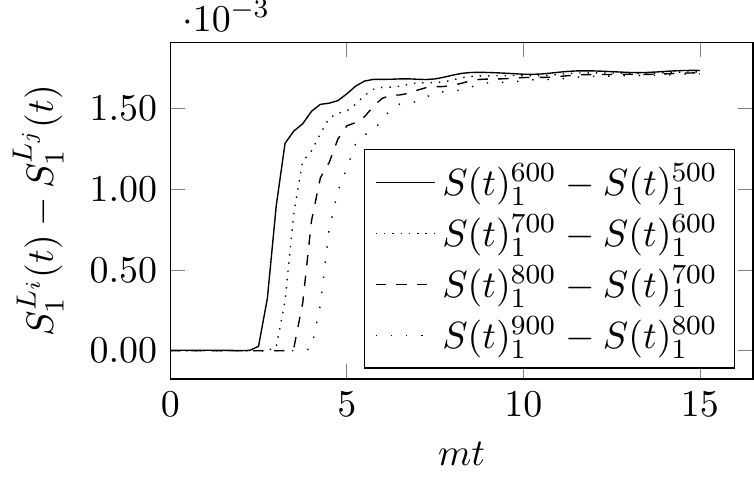} 

 \caption{\textit{Left:} The time evolution of the von Neumann entropy after the mass quench $m_0=0.0095\rightarrow m=0.01$ for various subsystem sizes with $a=1$. The symbols correspond to the
numerical evaluation of the correlation matrix, while the continuous line is  $\lim_{n\rightarrow 1} 2(C_{2,0}(t)+C_{0,2}(t))+1.1082$. For short times, before the saturation sets in, the points corresponding to different subsystem sizes overlap. After the saturation, all curves exhibit oscillations that persist for large times, and are well reproduced by the formula~\eqref{C2002} for $C_{2,0}(t)+C_{0,2}(t)$ up to a constant offset and a factor of two,  due to clustering of the branch point twist field two-point function. The different heights of the curves are due to the different subsystem finite sizes and the presence of a contribution to the entanglement entropy that is proportional to the subsystem size. \textit{Right:} Differences between the von Neumann entropies calculated at different subsystem sizes after the same quench. For large enough times all curves coincide, demonstrating that the oscillatory part of the entropies is independent of the subsystem size and the dependence on the subsystem size is (as expected) linear.}
 \label{fig:vN_sat}
\end{figure}

As was already pointed out in~\cite{EEquench} for a finite subsystem size the entanglement entropy saturates to a constant after a finite time. The saturation constant is linear in the subsystem size. The authors studied large quenches, where the leading behaviour is the linear growth, and there was no trace of oscillatory behaviour. The left panel of fig.~\ref{fig:vN_sat} shows the time evolution of the von Neumann entropy of different subsytems for a small quench with fixed lattice spacing. It is clear that  also after saturation the entropy continues to oscillate. The baseline of the oscillations saturates as well, but apart from this offset, the functional form is predicted well by the field theory formula~\eqref{C2002}, which in principle is not supposed to be valid for $t\gg L$.

For a comparison we plot the entanglement entropy together  with~\eqref{C2002} shifted by an arbitrary constant. In the right panel of fig.~\ref{fig:vN_sat} we plot the differences of the entanglement entropy, calculated at different subsystem sizes. After the saturation the curves are equally spaced, which shows that the oscillations do not depend on the subsystem size.

\begin{figure}[t]
     \centering
       \includegraphics[width=7.5cm]{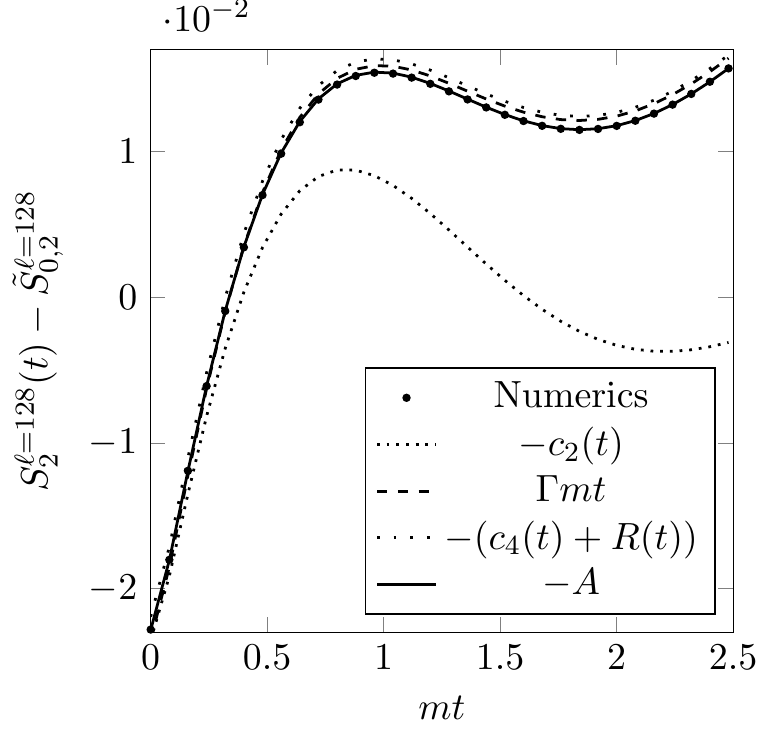} 
       \includegraphics[width=7.5cm]{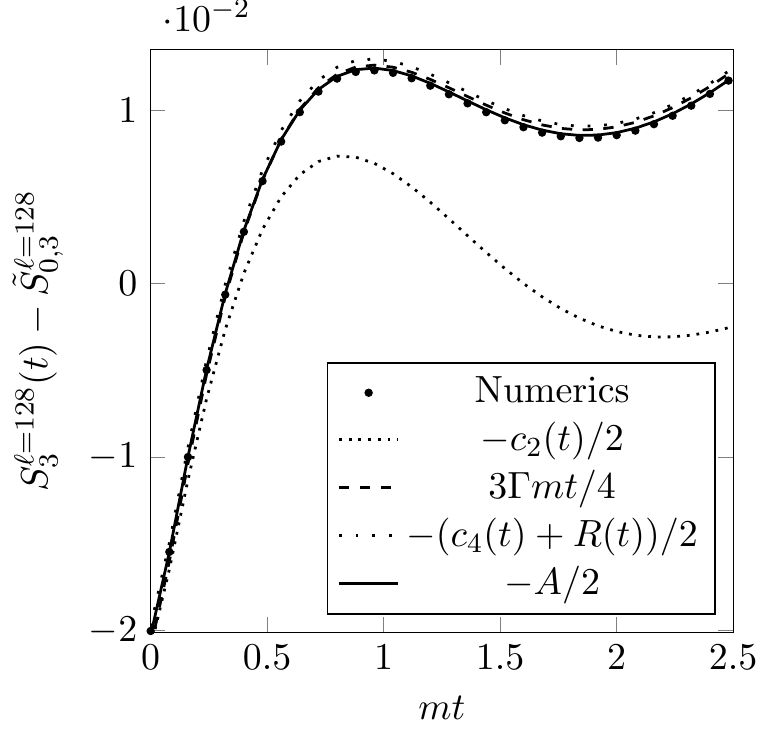} 
 \caption{The time evolution of the 2nd and 3rd R\'enyi entropies after the mass quench $m_0=0.048\rightarrow m=0.04$ for subsystem size $\ell=128$ extrapolated to $a=0$. The $\tilde{S}_{0,n}^l$ subtraction is the equilibrium entropy of the post-quench ground state, analogously to \eqref{pqVEV}. The curves exhibit both oscillations and linear growth. The dotted line corresponds to the contribution $c_2(t)=C_{2,0}(t)+C_{0,2}(t)$. Other curves incorporate the indicated contributions to~\eqref{final_Renyi2} one-by-one with  $c_{4}(t):=C^{(2)}_{40}(t)+C^{(2)}_{04}(t)$. The full prediction~\eqref{final_Renyi2} is in remarkably good agreement with the numerical data.}
 \label{fig:renyi_t}
\end{figure}

From fig.~\ref{fig:vN_sat} we can draw several conclusions:
\begin{itemize}
\item Before saturation, the values of the entanglement entropy are independent of the subsystem's size. This demonstrates the clustering of the two point function of twist fields.
\item The saturation times and saturation values are equally spaced for different subsystem sizes (with fixed difference in the size). This shows the $\propto L$ behaviour of the saturation values, which was already discussed in~\cite{EEquench,FC}.
\item The oscillations are present, independently of the linear growth and the saturation. After the saturation sets in, the shape of the oscillations is the same for different subsystem sizes. Moreover, they persist for large times and are well reproduced by the formula~\eqref{C2002} up to a constant offset, and a factor of two.  As already discussed, the factor of two is the result of the clustering.
\end{itemize}

\begin{figure}[t]
     \centering
\includegraphics[width=10cm]{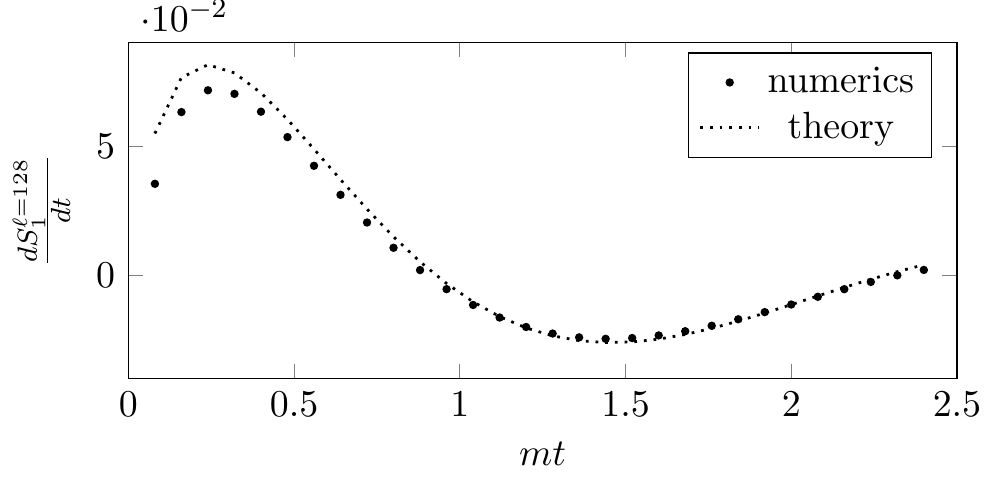} 
	\caption{The time derivative of the von Neumann entropy after the mass quench $m_0=0.048\rightarrow m=0.04$ for subsystem size $\ell=128$ extrapolated to $a=0$. The numerics was shifted by the value calculated in~\eqref{scalingEEint}.  In other words, the linear time growth has been subtracted to leave only the oscillatory part.}
	\label{fig:Sder}
\end{figure}

\subsubsection{Linear Growth and Oscillations}

To observe linear growth in time and test the field theory result~\eqref{final_Renyi2}, one needs larger subsystem sizes in order to prevent saturation within the time window. Fig.~\ref{fig:renyi_t} shows the time evolution of the R\'enyi entropies after a mass quench. The theoretical prediction~\eqref{final_Renyi2} is in remarkably good agreement with the numerical data. One can also see how the successive contributions of the different terms in~\eqref{final_Renyi2} improve accuracy. 

Within our field theoretical approach, as mentioned in section~\ref{sec:sum}, for the von Neumann entropy we can obtain only the oscillatory behaviour. Additional contributions are also expected to be present in this case. One can then take the time derivative in order to eliminate the time independent offset and subtract the value predicted by~\eqref{scalingEE2} to eliminate the linear growth, which, in turn, produces an offset in the time derivative. The results can be seen in fig.~\ref{fig:Sder}. The agreement with the field theory prediction is again very good except for the small $t$ region. Notice that  the corrections  coming from  $C^{(2)}_{40}(t)+C^{(2)}_{04}(t)$  and $R(t)$ can not be calculated at $n=1$ and, although subleading for large time, they might affect the small time behaviour.

\section{Conclusion}
\label{sec:conc}
In this paper we have presented an analytic derivation of the leading large-time post-quench dynamics of entanglement in the massive Ising field theory. We considered in particular a global quench resulting from a sudden change in the mass of the fermionic particle, from an initial value $m_0$ at time $t=0$ to a subsequent value $m$ for $t>0$.
 For the first time in a dynamical context for massive quantum field theories, we have employed the branch point twist field approach \cite{entropy} in our computations.
We have computed the R\'enyi entropies of a semi-infinite interval, which are proportional to the logarithm of the one-point  function of a branch point twist field in a replica quantum field theory.
In particular, the twist field one-point function has been computed exactly up to $O(K^2)$, in the post-quench quasi-particle expansion of the initial state by employing a regularization scheme for the infinite volume divergences discussed~in \cite{SE}.

Such an expansion can be also recast as a perturbative series in the quench parameter $\delta m:=m-m_0$, and is then exact up to $O(\delta m^2)$. At first order in the quench parameter $\delta m$ the result for the twist field one-point function can also be recovered from a perturbative expansion in the pre-quench quasi-particle basis by generalizing the approach introduced in~\cite{PQ1}.
We demonstrated, moreover, that crucial effects of the relaxation dynamics, such as linear growth of entanglement must manifest  as second order corrections in such a perturbative expansion. The main conclusions from the analytic results can be then summarized as follows:
\begin{itemize}
\item We showed the presence of a contribution to the R\'enyi entropies which grows linearly in time with slope $\frac{n \Gamma m}{2 (n-1)}$, where $\Gamma$ is up to $O(K^2)$ exactly the decay rate of the spin operator after a mass quench  found in \cite{SE}. 

\item The R\'enyi and von Neumann  entropies contain contributions which are oscillatory, with frequency of oscillation $2m$ and amplitude proportional to $(mt)^{-3/2}$ for large-time. Those contributions are of first order in $K$ and can be also obtained from a perturbative expansion in the quench parameter $\delta m$.
Our result implies that oscillations in the entanglement entropies are not produced by finite size-effects, as for instance stated in~\cite{FC}, but are rather inherent properties of those quantities. The same reasoning will apply to the constant shift of the R\'enyi entropies, which we also determined in section~\ref{massquench}.

\item We have provided a simple argument to show that, up to a constant normalization by the VEV of the twist field, the one-point function can be expressed  as the exponential of a Laurent polynomial in the variables $ (mt)^{\frac{p}{2}}$ for $p\leq 2$ and $p\neq 1$. We have shown this at $O(K^2)$ and expect to generalize this conclusion to higher orders in the future. Interestingly, the arguments leading to exponentiation of the one-point function also apply to the order parameter discussed in \cite{SE}. 
\end{itemize}

As expected, our field-theoretical results for the linear large-time behaviour of the R\'enyi entropies reproduce the scaling limit of the  formulae found in \cite{FC} up to $O(\delta m^2)$. The field-theoretical expansion, however, extends the lattice results to intermediate times and is confirmed with remarkable accuracy by numerical calculations directly in the scaling limit. Comparison between analytic and numerical results also shows that the two-point function of branch point twist fields after a global quench satisfies clustering, as previously observed for the spin field \cite{CEF,CEF2,SE}. Thus the entanglement entropies are proportional to the number of subsystem boundary points, just as in equilibrium situations.

 It would be interesting to use twist fields and the approaches discussed in section~\ref{sec:formfactor} and section~\ref{pert_theory} to consider other (small) global quenches. In particular, quenches that  drive the theory away from an integrable point or to a different interacting integrable model.  A particular case is the quench of the longitudinal magnetic field in the Ising spin chain while fixing the transverse field at its critical value $h=1$. In the scaling limit, this corresponds to a mass quench in the  so-called minimal $E_8$ Toda theory. The prediction is then that the entanglement entropies will oscillate with frequencies that are directly the quasi-particle masses of the $E_8$ field theory. An analogous phenomenon has been observed numerically in the Ising spin chain~\cite{Gabor1, Gabor2, Gabor3}, for a quench of the longitudinal magnetic field, but in the ferromagnetic  phase $h<1$. The presence of oscillations in the entanglement entropies and their slower (linear) growth in time were ascribed  to the confinement of the kinks.

Finally, it would be useful to develop a quasi-particle interpretation/derivation of the oscillatory contributions to entanglement. Even though our results are restricted to the Ising field theory, the emergence of such oscillations in the context of form factor expansions seems very natural. This suggests that it is a universal feature of quenches in gapped theories. 

A powerful unifying picture emerges from our work: the dynamics of entanglement and that of correlators of local fields after a global quench {\it{are not}} fundamentally distinct. Rather, the dynamics of entanglement is just the dynamics of correlators of a particular field, the branch point twist field. As a consequence, the large time linear growth of entanglement emerging from the quasi-particle picture of \cite{EEquench} is nothing but the exponential decay of correlators (in our case, the one-point function) at large time after the quench. Suggestively, out-of-equilibrium dynamics where no indication of linear growth of entanglement is observed, such as in the presence of confinement \cite{Gabor1, Gabor2, Gabor3}, could signal that certain local observables fail to relax exponentially fast at large times.
 The present work extends the seminal results of~\cite{quench} out-of-criticality and, for a very simple model, provides further evidence of the rich and interesting dynamics of correlators in out-of-equilibrium massive QFT. 

\paragraph{Acknowledgements:} We are indebted to Aldo Delfino, Fabian Essler and Dirk Schuricht for  discussions and helpful e-mail exchanges regarding their works \cite{PQ1, SE}, and to G\'abor Tak\'acs for sharing his insights into the wider interpretation and applicability of our results.

OCA and IMSZ are also grateful to Benjamin Doyon and Cecilia De Fazio for many useful discussions and feedback. They gratefully acknowledge support from EPSRC through the standard proposal {\it ``Entanglement Measures, Twist Fields, and Partition Functions in Quantum Field Theory"} under reference number EP/P006108/1 and from the International Institute of Physics in Natal (Brazil). This project was initiated during the Workshop on {\it Transport in Strongly Correlated Systems} held there in the summer of 2018. OCA's research was partly supported by an {\it Emmy Noether Visiting Fellowship} of the Perimeter Institute for Theoretical Physics.  Research at Perimeter Institute is supported by the Government of Canada through the Department of Innovation, Science and Economic Development and by the Province of Ontario through the Ministry of Research, Innovation and Science.

 ML is grateful to F\'abio Novaes for general discussions. ML and JV acknowlegde financial support from the Brazilian ministries MEC and MCTIC. 
\appendix

\section{Numerics in the Scaling Limit}
\label{app:a}
\subsection{Extrapolations and Extraction of the Central Charge and Scaling Dimensions}
\label{ccharge}
In this appendix we present further details on the numerical scaling limit discussed in section~\ref{sec:numerics}. The central charge and the operator scaling dimensions can be extracted from the evaluation of the restricted correlation function at $t=0$. We fix $m$ to a certain value in the scaling field theory, and change $a$ and $L$ in such a way that the physical subsystem size $\ell=aL$ is kept fixed. Then one can fit the lattice spacing dependence of the logarithm of the one-point function of the disorder operator and the R\'enyi entropies with the following functional forms
\beqa
\log \langle\mu(a)\rangle&\approx&A + B \log a + C a + D a^{2}\,,\\
S_{n}&\approx&A + B \log a + C a^{1/n} + D a^{2/n}\,.
\eeqa
For the R\'enyi entropies unusual corrections are present~\cite{CalCarPesUnusual}. For the disorder operator one assumes standard corrections, more on this operator can be found in section~\ref{scalim}.

The coefficient of the $\log a$ term corresponds to the scaling dimension of the operators or in the case of the von Neumann entropy the central charge of the UV CFT. We carried out the fit with the following parameters: $m=0.04,\,\ell=128,\,a=1/4,1/7,1/8\dots 1/20$. The results are summarized in Table~\ref{tab:scafits}.

\begin{table}[h!]
\centering
\begin{tabular}{|c|c|c|c|c|c|}
\hline
   & $c$ & $\Delta_{\mu}$ & $\Delta_{2}$& $\Delta_{3}$& $\Delta_{4}$ \\
\hline
 Theory & 0.5 & 0.125 & 0.0625 & 0.11111 & 0.15625 \\
\hline
 Fit & 0.50195(3) & 0.124969(4) & 0.06258(1) & 0.11002(9) &  0.1525(3) \\
\hline
\end{tabular}
\label{tab:scafits}
\caption{UV central charge and operator dimensions from the scaling limit extrapolations in the ground state with $m=0.04$ and $\ell=128$.}
\end{table}

The agreement for the central charge, the dimension of the disorder operator, and of the twist field with $n=2,3,4$ is excellent (see equation (\ref{stwist})). Note that the central charge can be extracted with better precision by calculating the von Neumann entropy at the critical point, with fixed lattice spacing and changing the number of sites in the subsystem, based on the logarithmic violation of the area law. In our case we extract the UV central charge away from the critical point, therefore we have less precision.

Using the above fits one can extrapolate to $a=0$. At different times, we used the same set of lattice spacings to carry out the extrapolations. Note that in our comparison we subtract the post-quench ground state entropy from the numerical results. The logarithmic singularity cancels from these differences, therefore we did not include the logarithm when extrapolating these quantities.

Note that going closer to the critical point would require more computational power since one has to increase the subsystem size correspondingly, therefore the size of the correlation matrix increases. The calculation of the intergrals~\eqref{eq:fg_def} gets also more difficult.
One also has to make sure, that the chosen subsystem size is large enough for the clustering of the two-point functions. For the quench studied in this paper we checked this using the saturation of the post-quench entropies. We found that for $a=1$ and $L\approx120$ the entropies are saturated up to $O(10^{-6})$, therefore we claim that our numerical results for the entropies have errors of this order.

\subsection{Scaling Limit and the Disorder Operator}
\label{scalim}

In \cite{CEF,CEF2} the decaying exponential characterizing the post-quench behaviour of the spin operator was found to be
\beq
\log\bra\sigma(t) \ket=t\int_{0}^\pi \frac{d\varphi}{\pi} |\epsilon'_{\varphi}| \log(\cos(\Phi_{\varphi}))\,,
\label{logs}
\eeq
therefore, in the scaling limit, see~\eqref{def_cos_sl}
\beq
\label{scal_mu_SH}
\lim_{\text{scal}}\log\bra\sigma(t) \ket= t \int_0^\infty \frac{dp~p}{\pi E_m(p)}\log(\zeta(p))\,.
\eeq
For small quenches, eq.~\eqref{scal_mu_SH} becomes
\beq
\label{scal_mu}
\log\bra\sigma(t) \ket= -\frac{t \, \delta m^2}{2 \pi m_0 }\int_0^\infty d\theta \, \frac{\tanh^3 \theta}{\cosh \theta}+O(\delta m^3)=-\frac{t \, \delta m^2}{3 \pi m_0 }+O(\delta m^3)\,,
\eeq
that agrees with the field theory result presented in~\cite{SE}, when expanded up to the second order in the quench parameter.

In~\cite{CEF2} the authors determine the time dependence of the order parameter by calculating the determinant of the correlation matrix. However, their definition of the correlation matrix is slightly different. In particular, in our notations, their T\"oplitz matrix starts with $\Pi_{-1}$ in the upper left corner. From~\cite{BarouchMcCoy} one can see that this can be absorbed into the redefinition of $h\rightarrow1/h$, realizing the Kramers--Wannier duality. Therefore calculating the determinant of~\eqref{eq:Cmatr} gives the square of the two-point function of the disorder operator, up to a constant:
\begin{equation}
 \mathrm{Det}\,\Gamma_L(t)\propto\left(\langle \mu(L,t)\mu(0,t)\rangle_{\mathrm{lattice}}\right)^2\propto \, a^{2\Delta_{\mu}}\, \left(\langle \mu(\ell=a L,t)\mu(0,t)\rangle_{\mathrm{field\,theor.}}\right)^2\,.
\end{equation}
Using the fitting procedure outlined in section~\ref{ccharge} we obtained $\Delta_{\mu}\approx0.12497$, which is very close to the theoretical value $\frac{1}{8}=0.125$.

\begin{figure}
\centering
\includegraphics[width=15cm]{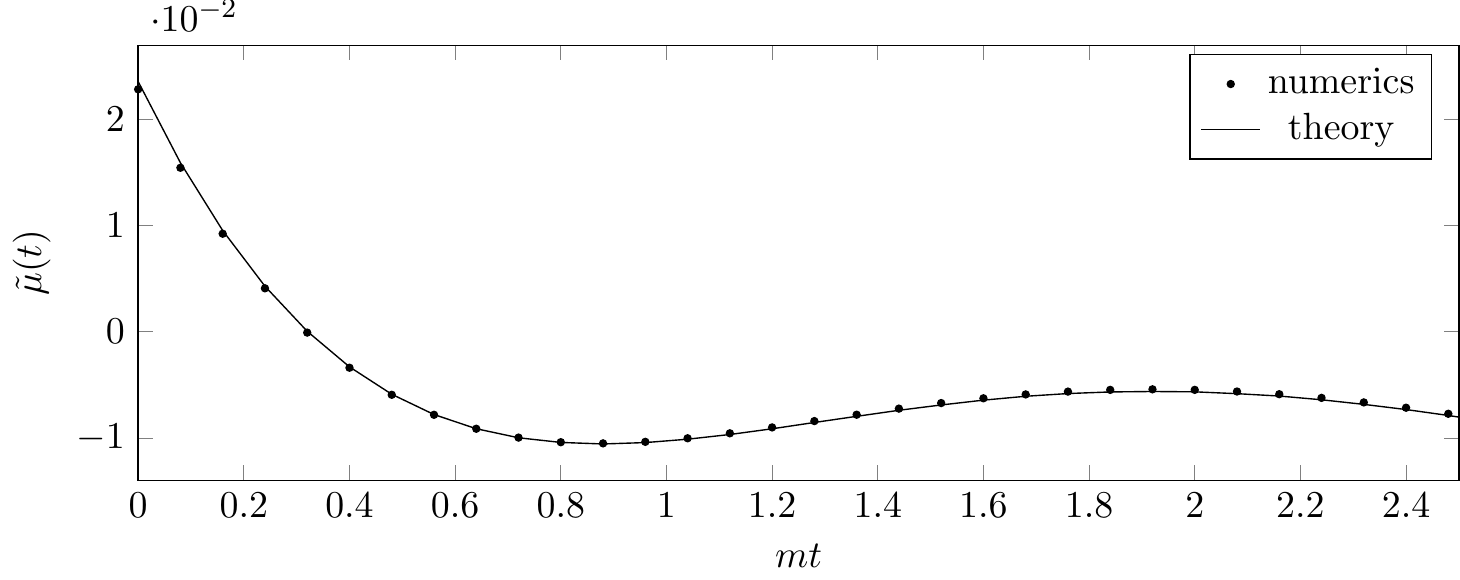} 
\caption{Time evolution of $\tilde{\mu}(t)$ compared to~\cite{SE} after quench $m_0=0.048\rightarrow m=0.04$ in the paramagnetic phase. The dots are the numerical results extrapolated to $a=0$, the line is the theoretical prediction for the oscillation and the linear growth of~\cite{SE}. The agreement is excellent, and it is clear that for smaller times one has to take into account the $1/t$ correction and there is no visible offset.}
\label{fig:mu}
\end{figure}

If the separation is large, the two-point function of $\mu$ clusters, just as for the order parameter in the ferromagnetic phase~\cite{CEF2}
\begin{equation}
 \label{eq:mucluster}
 \langle \mu(\ell,t)\mu(0,t)\rangle = (\langle \mu(0,t)\rangle)^2+O(\mathrm{e}^{-\ell m})\,.
\end{equation}
Therefore for large enough separations one can get access to the one-point function. It can be also seen that in the scaling limit $\tilde{\mu}(t)=\log \langle\tilde{0} |\mu(0,t)|\tilde{0}\rangle-\log\langle\tilde{0} |\mu(0,0)|\tilde{0}\rangle$ has a finite limit. Based on the Kramers--Wannier duality the formulas of~\cite{SE} for the order parameter in the ferromagnetic phase can be directly used to test the disorder operator in the paramagnetic phase. Such a comparison can be seen in fig.~\ref{fig:mu}. The agreement is excellent. Note that in the case of the order/disorder operator there is no offset at $O(K^2)$~\cite{SE}.

\end{document}